\documentclass[onecolumn, aps, nofootinbib, prd, preprint, floats, floatfix, amsmath, amssymb, superscriptaddress, preprintnumbers]{revtex4-1}

\usepackage[dvipsnames]{xcolor}
\usepackage{tikz}
\usetikzlibrary{positioning,calc,arrows.meta}
\usepackage{makecell}
\usepackage{graphicx}
\usepackage{subfigure}
\usepackage{dcolumn}
\usepackage{bm}
\usepackage{amssymb}
\usepackage[utf8]{inputenc}
\usepackage[OT1]{fontenc}
\usepackage{yfonts,amsmath,amsthm,amsfonts,amssymb,amscd, ulem}
\usepackage{enumerate}
\usepackage{fancyhdr}
\usepackage{mathtools}
\usepackage{mathrsfs}
\usepackage{cancel}
\usepackage{slashed}
\usepackage{bigints}
\usepackage[flushleft]{threeparttable}
\usepackage[colorlinks=true, citecolor=purple, linkcolor=blue]{hyperref}
\usepackage{url}
\usepackage{makecell,booktabs}
\usepackage{braket}
\usepackage{relsize}
\usepackage{multirow}
\usepackage{placeins}
\usepackage{verbatim}
\usepackage{txfonts}
\usepackage{slashed}
\usepackage{upgreek}
\usepackage{extarrows}
\usepackage{array}
\usepackage{appendix}
\usepackage[T1]{fontenc}
\usepackage{setspace}

\usepackage{tikz} 
\usetikzlibrary{shapes,arrows,positioning,automata,backgrounds,calc,er,patterns}
\usepackage{tikz-feynman}
\tikzfeynmanset{compat=1.0.0}
\usetikzlibrary{shapes.misc}
\tikzset{cross/.style={cross out, draw=black, fill=none, minimum size=2*(#1-\pgflinewidth), inner sep=0pt, outer sep=0pt}, cross/.default={2pt}}
\usetikzlibrary{shapes.geometric}
\usepackage{orcidlink}

\renewcommand{\arraystretch}{0.6}

\begin{document}

\title{Down-Type Jet Identification in Fully Hadronic $t\bar t$ Events}

\author{Jinhui Guo}
\email{guojh23@buaa.edu.cn}
\affiliation{School of Physics, Beihang University, Beijing 100191, China}

\author{Chenhao Peng }
\email{chenhaopeng@stu.pku.edu.cn}
\affiliation{School of Physics and State Key Laboratory of Nuclear Physics and Technology, Peking University, Beijing 100871, China}

\preprint{CPTNP-2025-026}

\begin{abstract}
Fully hadronic $t\bar t$ events carry the largest branching fraction and provide the down-type quark as a near-maximal spin analyzer ($\beta_d\simeq1$), yet they are rarely used for spin-correlation and entanglement studies because the down-type jets must be extracted from a difficult jet-to-parton reconstruction in the presence of a large QCD multijet background.  
We develop a two-stage machine-learning reconstruction for this channel: a GNN+Transformer network assigns all reconstructed jets into a legal six-jet top-pair candidate, and a second hybrid classifier resolves the four down/up hypotheses inside the two hadronic $W$ decays.  The second stage combines an assignment scorer with an auxiliary conditional diffusion head; we find that the diffusion assignment score can rank hypotheses only when trained with a margin ranking objective, while standard denoising objectives leave it at the random baseline, and the two objectives are largely decoupled.  
We evaluate the method against a calibrated QCD background and an unmatched $t\bar t$ component using both reconstruction and spin observables.  Compared with a ST1 only benchmark with random down/up labels, the learned ST2 classifier improves both the effective spin-analysis factor and the purity-weighted six-parton exact fraction at fixed event selection.  
A direct check of the standard spin-correlation coefficient $D$ further gives an ML-reconstructed value compatible with the truth-level value within the resampled spread.  These results show that down/up identification can retain useful spin-correlation information in the all-hadronic channel.
\end{abstract}

\maketitle

\tableofcontents

\section{INTRODUCTION}

The top quark decays before hadronization, so the angular distributions of its decay products retain information about the production spin density matrix~\cite{Bernreuther:2008ju,Mahlon:2010gw,Bernreuther:2015yna}.  For a polarized top quark, $(1/\Gamma)d\Gamma/d\cos\theta_i=(1+\beta_iP\cos\theta_i)/2$, where $\beta_i$ is the spin-analyzing power of decay product $i$.  In $t\bar t$ events, pairs of such directions probe the spin-correlation matrix and entanglement-sensitive combinations such as the trace observable $D$.  This connection motivated collider studies of quantum tomography, Bell inequalities, and entanglement~\cite{Afik:2020onf,Fabbrichesi:2021npl,Severi:2021cnj,Aguilar-Saavedra:2022uye,Afik:2022kwm}, followed by the ATLAS and CMS dilepton observations~\cite{ATLAS:2023fsd,CMS:2024pts}.  Semileptonic decays have also been proposed for these measurements~\cite{Han:2023fci}, and CMS has measured the full spin-correlation matrix and observed entanglement at high $m_{t\bar t}$ in lepton+jets events using a neural-network reconstruction of the hadronic branch~\cite{CMS:2024zkc}.  These results make hadronic spin analyzers experimentally relevant beyond the dilepton channel.

The fully hadronic channel is statistically attractive: ${\cal B}(t\bar t\to b\bar b+4q)\simeq45\%$, about four times the inclusive dilepton rate and ten times the $e/\mu$ dilepton rate~\cite{ParticleDataGroup:2024cfk}, and the down-type fermion from $W\to q\bar q'$ is a near-maximal analyzer, with $|\beta_d|=1$ at leading order~\cite{Bernreuther:2008ju,Brandenburg:2002xr}.  Event yield alone, however, does not determine the usable spin information: the measured correlation is diluted by the effective analyzing powers of both reconstructed directions.  A resolved event requires two $b$ jets and two $W$-daughter pairs to be assigned to the correct top branches, followed by identification of the down-type member of each pair, in the presence of additional radiation and a large QCD multijet background.  Wrong assignments or down/up interchanges can therefore substantially offset the branching-fraction advantage.

Hadronic polarimetry has progressively reduced the light-jet ambiguity.  Energy-ordering and subjet-based analyzers were developed in Refs.~\cite{Brandenburg:2002xr,Krohn:2009wm}, while the optimal kinematic polarimeter weights the two light-jet directions by their conditional down-type probabilities and reaches an analyzing power of about $0.64$ at leading order~\cite{Tweedie:2014yda}.  Constituent-level charge and geometry provide additional flavour information~\cite{Fraser:2018ieu}; recent GNN and analytic polarimeters combine kinematics, charge, and multiplicity to improve upon the kinematic construction~\cite{Dong:2024xsg,Dong:2024xsb}.  The recent down/up studies in Refs.~\cite{Dong:2024xsg,Dong:2024xsb} reconstruct a boosted hadronic top in semileptonic $t\bar t$ and use parton--subjet matching to define the down/up task.  They establish that constituent-level information improves the analyzer, but do not address the event-wide combinatorics and QCD background of the resolved fully hadronic channel.

Event-level reconstruction has developed in parallel~\cite{Barman:2024wfx}, from likelihood fits and permutation-scoring networks~\cite{Erdmann:2013rxa,Erdmann:2019evj} to attention, graph, and hypergraph architectures for high-multiplicity final states~\cite{Lee:2020qil,Fenton:2020woz,Shmakov:2021qdz,Ehrke:2023cpn,Birch-Sykes:2024gij}.  Attention-based methods avoid explicit factorial enumeration and encode decay symmetries, while graph-based methods represent the intermediate top and $W$ structure; direct four-momentum regression and matrix-element-guided assignments provide complementary strategies~\cite{Qiu:2022xvr,Dillon:2025dxr}.  In the published SaJa, SPANet, Topograph, and HyPER all-hadronic benchmarks~\cite{Lee:2020qil,Fenton:2020woz,Ehrke:2023cpn,Birch-Sykes:2024gij}, the two daughters of each $W$ are treated as exchange-equivalent, while direct momentum regression does not return their ordered identities.  This is sufficient for the parent momentum, but not for spin analysis because the down- and up-type directions have different analyzing powers.  The CMS lepton+jets analysis orders these jets on its single hadronic branch~\cite{CMS:2024zkc}; the unresolved fully hadronic task requires this ambiguity to be resolved on both branches without a charged-lepton anchor.

We address this task with a two-stage machine-learning reconstruction.  A GNN+Transformer network first uses jet-level and constituent-level information to form legal six-jet $t\bar t$ candidates.  A second classifier then ranks the four joint down/up hypotheses, with a conditional diffusion head testing whether a denoising score can also rank discrete assignments.  We evaluate the method at simulation level with calibrated QCD and unmatched $t\bar t$ components, using both reconstruction and spin observables.  Standard denoising objectives leave the diffusion assignment score near the random baseline, whereas a margin-ranking objective makes it discriminative and supplies auxiliary selection variables.  At $Q_{\rm ST1}^{\rm avg}\ge-0.0889$, the learned assignment and a random-label benchmark have the same $N_{\rm bkg}/N_{\rm sig}=0.631$, while $\eta_\kappa$ increases from $0.152$ to $0.255$ and the purity-weighted six-parton exact fraction from $0.139$ to $0.218$.  The reconstructed $D$ is also statistically compatible with the truth-level result within the current resampling uncertainty, directly testing whether the recovered directions retain the target spin correlation.

The remainder of this paper is organized as follows. In Sec.~\ref{sec:datasets}, we describe the signal and background samples, truth matching, and inputs.  Section~\ref{sec:network} presents the two-stage network and training objectives.  In Sec.~\ref{sec:results}, we study diffusion ranking, reconstruction quality, background rejection, and retained spin information.  Section~\ref{sec:conclusion} concludes.

\section{Signal and Background Datasets}\label{sec:datasets}
This section defines the datasets used in the fully hadronic $t\bar t$ analysis. We first describe the simulated $t\bar t$ sample, the truth matching procedure, and the resulting matched and unmatched samples. We then introduce the QCD multijet background and its ATLAS-based normalization.

\subsection{Signal Datasets}\label{sec:sig-data}
We simulate the production of all-hadronic $t\bar t$ pairs in $pp$ collisions at $\sqrt{s}=13~\mathrm{TeV}$ using {\tt MadGraph5\_aMC@NLO} \cite{Alwall:2014hca}. 
The top quarks are decayed with {\tt MadSpin} \cite{Artoisenet:2012st} enforcing $t\to bW$ and $W\to q\bar q'$, and then the resulting events at parton level are passed to {\tt PYTHIA} \cite{Bierlich:2022pfr} for parton showering and hadronization, followed by {\tt DELPHES} \cite{deFavereau:2013fsa} for fast detector simulation and particle-flow reconstruction. 
Apart from setting the FastJetFinder and GenJetFinder clustering radius to $R=0.3$, the default CMS Delphes card settings are retained throughout this analysis \cite{Lee:2020qil}.

Particle-flow (PF) objects in Delphes, namely EFlowTrack, EFlowPhoton, EFlowNeutralHadron, and EFlowMuon candidates when available, are then clustered with the anti-$k_T$ algorithm with radius parameter $R=0.3$ to define the analysis jets \cite{deFavereau:2013fsa}. The use of particle-flow objects enables us to preserve the constituent-level information of the jets, which is used as input to the subsequent machine-learning analysis.
Following the selection triggers used in CMS $t\bar{t}$ pair analysis in Ref. \cite{CMS:2018tye}, all the jets from $t\bar{t}$ decays are required to satisfy $p_T>30~\mathrm{GeV}$ and $|\eta|<2.4$, and events are selected if they contain at least six jets with $p_T>40~\mathrm{GeV}$, at least one $b$-tagged jet, and $H_T=\sum p_T>450~\mathrm{GeV}$. 
For the analysis jets reclustered from PF candidates, the Delphes $b$-tagging information is propagated by matching each reclustered jet to the nearest Delphes jet within $\Delta R<0.25$. Events with any unmatched reclustered jet are rejected as part of the preselection.

For preselected events, truth labels are assigned by matching reconstructed jets to the six generator level partons from the top decay chain, following Ref. \cite{Lee:2020qil}.
For the labeling, we first identify the six partons from the top decay chain, including the two $b$ quarks and the two up-type and two down-type light partons from the $W$ decays, using generator-level mother-daughter links. The selected jets are then geometrically matched to these truth partons with a one to one constraint, requiring $\Delta R(\mathrm{jet},\mathrm{parton})<0.3$~\cite{Lee:2020qil}. The matching is performed greedily by assigning pairs in ascending order of $\Delta R$, so that each jet and each parton is used at most once.

Events with all six top decay partons matched to distinct jets are kept as the {\it fully matched} $t\bar t$ signal sample, while selected events failing this full matching requirement are stored as the {\it unmatched} sample. In total, $N_{\mathrm{tot}}$ events were generated, $N_{\mathrm{sel}}$ events passed the preselection, and $N_{\mathrm{sig}}$ events were fully matched. The remaining $N_{\mathrm{unm}}=N_{\mathrm{sel}}-N_{\mathrm{sig}}$ events cannot be fully reconstructed. In this analysis, 20 million events are generated, $6.27\%$ of the generated events passed the preselection, and $1.27\%$ of the generated events passed the preselection and were fully matched to all six top-decay partons, which gives 253009 fully matched events. The preselection unmatched-to-signal ratio is
\begin{equation}\label{eq:Rpre}
    R_{\rm unm/sig}^{\rm pre}
    =
    \frac{N_{\rm unm}^{\rm pre}}{N_{\rm sig}^{\rm pre}}
    =
    \frac{1{,}000{,}617}{253{,}009}
    \simeq 3.95 .
\end{equation}

Following Refs.~\cite{Lee:2020qil,Dong:2024xsg}, the ML input for each selected event contains jet-level and constituent-level information. For each reconstructed jet, we store a 14-dimensional jet feature vector and up to 40 PF constituents. The jet feature vector is defined as
\begin{equation}\label{eq:14-d-feature}
\mathbf{x}_j=\Big\{
p_{T,j},\,\eta_j,\,\sin\phi_j,\,\cos\phi_j,\,
\frac{p_{T,j}}{H_T},\,\mathrm{btag}_j,\,
\mathrm{pTD}_j,\,s_{1,j},\,s_{2,j},\,
f_{\mathrm{ch},j},\,f_{\gamma,j},\,f_{\mathrm{neu},j},\,f_{e,j},\,f_{\mu,j}
\Big\},
\end{equation}
where $H_T=\sum_j p_{T,j}$ is the total transverse momentum of the jet of each matched event, $\mathrm{btag}_j$ denotes the $b$-tagging information assigned to this jet, $\mathrm{pTD}_j=
\sqrt{\sum_{m\in\mathcal{J}_j} p_{T,m}^2}/
\sum_{m\in\mathcal{J}_j} p_{T,m}$ characterizes the concentration of transverse momentum among the PF constituents $\mathcal{J}_j$ of jet $j$, and larger $\mathrm{pTD}_j$ corresponds to jets whose transverse momentum is carried by fewer, harder constituents. 
The shape variables $s_{1,j}$ and $s_{2,j}$ describe the angular widths of jet $j$ as measured from its PF constituents. They are defined as the square roots of the two eigenvalues of the $p_T$ weighted covariance matrix $C_j=\frac{1}{\sum_{m\in\mathcal{J}_j}p_{T,m}}\sum_{m\in\mathcal{J}_j}p_{T,m}\left(\begin{smallmatrix}\Delta\eta_m^2 & \Delta\eta_m\Delta\phi_m\\ \Delta\eta_m\Delta\phi_m & \Delta\phi_m^2\end{smallmatrix}\right)$, ordered as $s_{1,j}\ge s_{2,j}$. Here $\Delta\eta_m=\eta_m-\eta_j$ and $\Delta\phi_m=\mathrm{wrap}(\phi_m-\phi_j)$ for each constituent $m\in\mathcal{J}_j$, where $(\eta_j,\phi_j)$ is the jet axis and $\mathrm{wrap}$ maps the azimuthal difference into $(-\pi,\pi]$. Finally, the particle-type fractions, $f_{\alpha,j}$ are defined as transverse-momentum fractions,
\begin{equation}
f_{\alpha,j}=
\frac{
\sum_{m\in \mathcal{J}_j,\,\mathrm{type}(m)=\alpha} p_{T,m}
}{
\sum_{m\in \mathcal{J}_j} p_{T,m}
},
\qquad
\alpha\in\{\mathrm{ch},\gamma,\mathrm{neu},e,\mu\},
\end{equation}
where the labels $\mathrm{ch}$, $\gamma$, $\mathrm{neu}$, $e$, and $\mu$ correspond to charged hadrons, photons, neutral hadrons, electrons, and muons, respectively. In the stored PF-candidate representation, electrons are identified by $\mathrm{pid}=\pm 11$, muons by $\mathrm{pid}=\pm 13$, photons by $\mathrm{pid}=22$, and the remaining constituents are classified as charged or neutral hadrons according to their charge.

For the {\bf constituent-level input}, each PF constituent $m\in\mathcal{J}_j$ of jet $j$ is represented by the 10-dimensional vector
\begin{equation}\label{eq:10-d-constituent}
 c_{m,j} =
 (\Delta\eta,\Delta\phi,p_T,E,Q,\mathbf{1}_e,\mathbf{1}_\mu,
    \mathbf{1}_\gamma,\mathbf{1}_{\mathrm{ch}},
    \mathbf{1}_{\mathrm{nh}}),
\end{equation}
where $\Delta\eta=\eta_m-\eta_j$ and $\Delta\phi=\mathrm{wrap}(\phi_m-\phi_j)$ are measured relative to the jet axis, $p_T$, $E$, and $Q$ are the transverse momentum, energy, and electric charge of the PF constituent $m$, and the last five entries form a one hot encoding of the PF constituent type: electron, muon, photon, charged hadron, or neutral hadron. For example, an electron constituent has $(\mathbf{1}_e,\mathbf{1}_\mu,\mathbf{1}_\gamma,\mathbf{1}_{\mathrm{ch}},\mathbf{1}_{\mathrm{nh}})=(1,0,0,0,0)$, whereas a charged-hadron constituent has $(0,0,0,1,0)$.

The resulting per-event inputs have variable numbers of jets and variable numbers of PF constituents per jet. For the ML inputs, we pad the jet collection to $J=17$ jets per event and the constituent collection to $N_{\rm const}^{\max}=40$ constituents per jet, setting the missing entries to zero.

\subsection{Background Datasets}\label{sec:background}
QCD multijet production is the dominant non-$t\bar{t}$ background for fully hadronic $t\bar{t}$ signals at a $pp$ collider. Using {\tt MadGraph5\_aMC@NLO}, we generate inclusive QCD multijets at $\sqrt{s}=13~\mathrm{TeV}$, with two to four final-state partons, binned in $H_T$ across $[200,300,500,700,1000,1500,2000,\infty)~\mathrm{GeV}$ to ensure sufficient high-$H_T$ statistics. Showering, hadronization, detector simulation, preselection, and jet/constituent feature extraction (14 jet-level + 10 constituent-level features per jet) follow the same pipeline as the previously described signal samples.

For the QCD multijet normalisation, we anchor the analysis to the ATLAS all-hadronic measurement~\cite{ATLAS:2020ccu}, supplemented by the cut flow details in Ref.~\cite{Poggi:2021thesis}. We define an ATLAS-like baseline, denoted by $B_{\rm ATL}$: the six leading selected jets must have $p_T>55~{\rm GeV}$, exactly two jets must pass the b-tag requirement, and a valid all-hadronic assignment with $\chi^2_{\min}<10$ must be found. From the corresponding ATLAS cut flow yields, we use $R^{B_{\rm ATL}}_{{\rm QCD}/t\bar t}\simeq2.99$ as the QCD-to-all-hadronic-$t\bar t$ calibration factor. The simulated QCD sample is then rescaled by this factor when evaluating the reconstruction performance. The $\chi^2_{\min}$ score entering $B_{\rm ATL}$ is computed independently of the network logits by assigning the two b-tagged jets to the two top candidates, enumerating four selected non-b jets as the $W$-daughter candidates, and minimizing
\begin{equation}
    \chi^2
    =\frac{(m_{t,1}-m_{t,2})^2}{2\sigma_t^2}
    + \frac{(m_{W,1}-m_W)^2}{\sigma_W^2}
    + \frac{(m_{W,2}-m_W)^2}{\sigma_W^2},
\end{equation}
where $m_{t,i}$ ($m_{W,i}$) is the reconstructed top quark ($W$ boson) mass, and the reference values of the $W$ boson mass, $W$ mass and top quark mass resolutions ($m_W,\sigma_t,\sigma_W$) are listed in Table~\ref{tab:hyperparameters}.

The QCD inference sample contains $7\times 10^{5}$ generated events, with $10^{5}$ events in each $H_T$ bin. After preselection and the $B_{\rm ATL}$ baseline, 44,455 and 1,124 events remain, respectively. The physical normalization is fixed by the ATLAS-calibrated factor $R^{B_{\rm ATL}}_{{\rm QCD}/t\bar t}$ introduced above. 
In addition to QCD multijets, unmatched $t\bar t$ events are also treated as a separate background component. For our generated 20 million signal events, preselection yields 1,000,617 unmatched and 253,009 matched fully hadronic $t\bar t$ events. After the $B_{\rm ATL}$ baseline, 39,331 unmatched and 33,572 matched events survive, giving
$R_{\rm unm/sig}^{B_{\rm ATL}}\simeq 1.17$.

After the preselection and the $B_{\rm ATL}$ baseline, the calibrated yield ratio of QCD plus unmatched $t\bar t$ background to the matched fully hadronic $t\bar t$ signal is
\begin{equation}
\begin{aligned}
    R_{\rm bkg/sig}^{B_{\rm ATL}}&=\frac{N_{\rm QCD}^{B_{\rm ATL}}+N_{\rm unm}^{B_{\rm ATL}}}{N_{\rm sig}^{B_{\rm ATL}}}\\
    &=R^{B_{\rm ATL}}_{{\rm QCD}/t\bar t}\cdot(1+R_{\rm unm/sig}^{B_{\rm ATL}}) + R_{\rm unm/sig}^{B_{\rm ATL}}\\
    &=7.67
\end{aligned}
\end{equation}
where $N_{\rm t\bar{t}}^{B_{\rm ATL}}=N_{\rm unm}^{B_{\rm ATL}}+N_{\rm sig}^{B_{\rm ATL}}$ has been used. In Table \ref{tab:cut-eff}, the relevant event numbers after preselection, the $B_{\rm ATL}$ baseline, and matching procedures are organized and listed.
\begin{table}[htbp]
    \centering
    \begin{tabular}{c | c | c | c | c | c | c }
    \hline
        Process & Generated & Preselection & Preselection unm./sig. & $B_{\rm ATL}$ & $B_{\rm ATL}$ unm./sig. & $R_{\rm unm/sig}^{B_{\rm ATL}}$  \\
        \hline
       $t{\bar t}$ (full-hadronic) & $2\times 10^{7}$ & 1,253,626 & 1,000,617/253,009 & 72,903 & 39,331/33,572 & 1.17 \\
       \hline
       QCD Multijets & $7\times 10^5$ & 44,455 & -- & 1,124 & -- & -- \\
    \hline
    \end{tabular}
    \caption{Event yields for the fully hadronic top-quark pair production and QCD multijets after the preselection, $B_{\rm ATL}$ baseline, and matching procedures.}
    \label{tab:cut-eff}
\end{table}

These samples define the inputs for the two-stage machine-learning reconstruction of fully hadronic $t\bar{t}$ events. Table \ref{tab:datasets} summarizes the sample sizes used for training and inference, together with the event counts passing the $B_{\rm ATL}$ baseline; the physical QCD normalization is applied through the calibration above.
\begin{table}[htbp]
\centering
\begin{tabular}{ccc}
\hline
Dataset & Event Number & After $B_{\rm ATL}$ \\
\hline
Training matched signal & 202,407 & 26,802 \\
Validation matched signal & 25,300 & 3,432 \\
Test matched signal & 25,302 & 3,338 \\
Unmatched $t\bar t$ inference sample & 1,000,617 & 39,331 \\
QCD multijet inference sample & 44,455 & 1,124 \\
\hline
\end{tabular}
\caption{Event numbers for the training and inference datasets before and after the $B_{\rm ATL}$ baseline.}
\label{tab:datasets}
\end{table}

\section{Hybrid Network: GNN+Transformer+Diffusion}
\label{sec:network}
Using the constructed datasets, we train and evaluate the proposed hybrid network. This section first introduces the overall architecture and the two classifier stages: ST1 for coarse jet assignment and ST2 for down/up assignment. We then define the loss functions, assignment scores, and cut variables used in the analysis, and finally summarize the hyperparameters and training procedure.

\textbf{Remark}: Given the large number of neural network hyperparameters, detailed descriptions for each would lead to lengthy text. For conciseness, all hyperparameters, including their notations, values and physical interpretations, are summarized in Table \ref{tab:hyperparameters}, and we do not elaborate on them individually in the main text.

\subsection{Network Architecture}
Inspired by Refs. \cite{Lee:2020qil,Dong:2024xsg}, we use a two step hybrid architecture with GNN, Transformer and Diffusion components for fully hadronic $t\bar{t}$ reconstruction and down type quark classification. We first give an overview of the two classifier stages, and then describe their assignment procedures in detail.

The first classifier, denoted as the Step-1 classifier (ST1), acts on all reconstructed jets in an event and outputs per jet logits over the five coarse labels
\begin{equation}
    y_j^{(5)} \in \{\mathrm{other}, b_1, W_1, b_2, W_2\}.
\end{equation}
For each event, the ST1 logits are used to build legal six-jet top-pair assignment candidates. The selected candidate is then passed to the Step-2 classifier (ST2), which resolves the down/up ambiguity inside the two hadronic $W$ decays. For convenience, the input order for ST2 is organized as
\begin{equation}\label{eq:tt-candidate}
    (b_1,q_{1a},q_{1b},b_2,q_{2a},q_{2b}),
\end{equation}
where $q_{ia}$ and $q_{ib}$ belong to $W_i$ for $i=1,2$. 
The target of ST2 is a two-bit label $\mathbf{Z}=(Z_1,Z_2)\in\{0,1\}^2$ for each $q_{ia}$--$q_{ib}$ combination: $Z_i=0$ means $q_{ia}$ is the down-type jet in branch $i$, and $Z_i=1$ means $q_{ib}$ is the down-type jet.

\begin{figure}[t]
    \centering
    \includegraphics[width=0.98\linewidth,trim=15pt 24pt 115pt 6pt,clip]{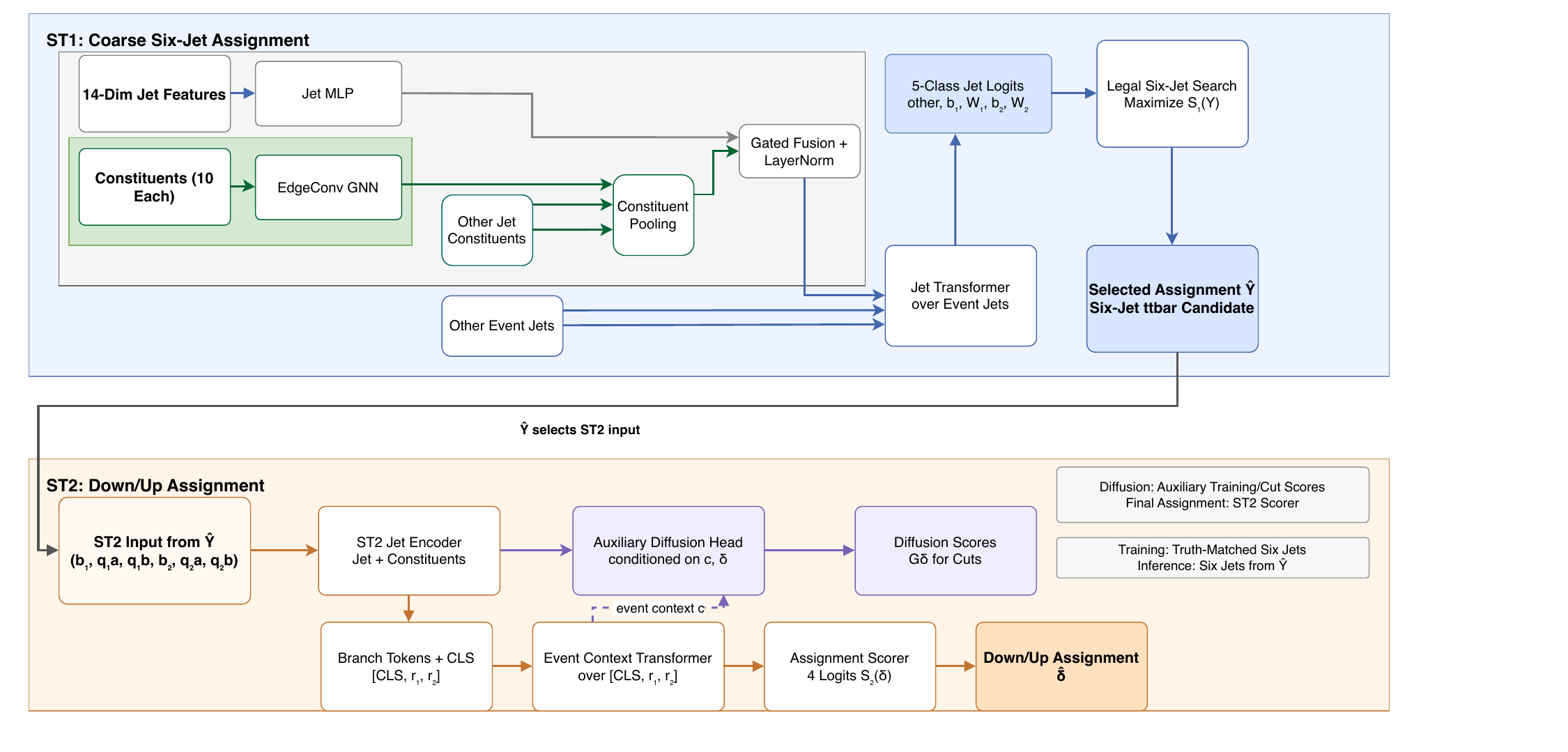}
    \caption{Network architecture. ST1 uses jet and constituent inputs to build the best legal assignment of six jets to a $t\bar t$ candidate; this ordered candidate is then passed to ST2. ST2 resolves the down/up assignment and includes an auxiliary diffusion branch for training and cut variables.}
    \label{fig:two-stage-network}
\end{figure}

Fig. \ref{fig:two-stage-network} presents the two-stage network architecture. The ST1 inputs are the preprocessed 14-dimensional jet-level features and 10-dimensional constituent-level features defined in Eqs. \eqref{eq:14-d-feature} and \eqref{eq:10-d-constituent}. Each jet-level feature is independently min--max scaled to the range $[0,1]$ using the minimum and maximum values fitted on the training set. For constituent-level features, $p_T$ and $E$ first undergo logarithmic transformation, and all constituent features are then standardized using the mean and standard deviation derived from valid constituents in the training set. The fitted preprocessing parameters are then kept fixed for validation, test, and inference samples.

ST2 takes six preselected jet candidates along with their complete jet-level and constituent-level features as inputs. 
During training, ST2 is supplied with truth matched six jet candidates; during inference, the six jets are taken from the legal assignment selected by ST1.

In summary, ST1 uses jet-level and constituent-level information to select the highest-scoring legal six-jet $t\bar{t}$ candidate, which is then passed to ST2 for down/up assignment. The diffusion module provides an auxiliary training signal in the second stage.

\subsubsection{Step-1 Classifier}
The ST1 classifier maps the per-event jet collection to per-jet five-class logits over $\{\mathrm{other}, b_1, W_1, b_2, W_2\}$. Each event is padded to at most $J=17$ jets, and each jet carries the 14-dim feature vector $x_j$ together with up to 40 constituents. The ST1 path shown in Fig. \ref{fig:two-stage-network} consists of a jet MLP encoder, a constituent GNN encoder, a gated fusion, a Transformer over jets, a per-jet classification head, and a final legal assignment.

For each valid jet $j$, the 14-dimensional jet vector is encoded by a two-layer multilayer perceptron (MLP),
\begin{equation}\label{eq:u-j}
    u_j = E_{\mathrm{jet}}(x_j) \in \mathbb{R}^d,
\end{equation}
where $d$ denotes the common token width used throughout both classifiers. The width for hidden layers is $h_{\rm w}d$, and each linear layer is followed by a ReLU activation.

In parallel with the jet feature encoder, the constituent encoder is a two block EdgeConv based constituent encoder, following the ParticleNet architecture in Ref. \cite{Qu:2019gqs}. For each jet, we construct a $K_{\rm GNN}$-nearest-neighbour graph whose nodes are the valid PF constituents of that jet, using their coordinates in the $(\Delta\eta,\Delta\phi)$ plane. The set of valid neighbors of constituent $m$ in jet $j$ is denoted by $\mathcal{N}_{jm}$. For block $\ell$, we have
\begin{equation}
    h_{jm}^{(\ell+1)}
    =
    \frac{1}{K_{\rm GNN}}
    \sum_{n\in\mathcal{N}_{jm}}
    \Phi_\ell\!\left(
    \left[
    h_{jm}^{(\ell)},\,
    h_{jn}^{(\ell)}-h_{jm}^{(\ell)}
    \right]
    \right)
    + R_\ell h_{jm}^{(\ell)} ,
\end{equation}
where $h_{jm}^{(0)} $ is the preprocessed feature vector of constituent $m$ in jet $j$, as shown in Eq. \eqref{eq:10-d-constituent}. The edge network $\Phi_\ell$ is a three-layer bias-free MLP, where each linear layer is followed by LayerNorm \cite{Ba:2016jcy} and ReLU activation functions \cite{Nair2010RectifiedLU}. The residual map $R_\ell$ is a bias-free linear projection followed by LayerNorm. The first EdgeConv block maps $d_{\rm const}\to c_1$, and the second maps $c_1\to c_2$.
Then, the output constituent features are pooled over the valid constituent set $\mathcal{V}_j$ of jet $j$,
\begin{equation}
    \overline{h}_j^{(2)} =\frac{1}{|\mathcal{V}_j|}\sum_{m\in\mathcal{V}_j} h_{jm}^{(2)} .
\end{equation}
A final projection $E_{\mathrm{cnt}}$ maps $\overline{h}_j^{(2)}$ to the jet-level constituent token with the same dimension as $u_j$ in Eq. \eqref{eq:u-j},
\begin{equation}
    v_j=E_{\mathrm{cnt}}\!\left(\overline{h}_j^{(2)}\right)
    \in \mathbb{R}^d .
\end{equation}
The jet-level token $u_j$ and constituent-level token $v_j$ are fused as
\begin{equation}
    z_j
    =
    F_{\mathrm{fuse}}\!\left(
    \begin{bmatrix}
        u_j\\
        \sigma(g)v_j
    \end{bmatrix}
    \right),
    \qquad
    F_{\mathrm{fuse}}(y)=\mathrm{LayerNorm}\!\left(W_f y\right),
\end{equation}
where $W_f\in\mathbb{R}^{d\times 2d}$ is a bias-free linear projection, and $\sigma(g)=\frac{1}{1+\exp(-g)}$ is the Sigmoid function with the learnable scalar gate $g$ initialized to $g_0$.

The fused jet tokens are then passed through an $N_{\rm layer}$-layer transformer encoder $T_1$ with $N_{\rm head}$ attention heads and feed-forward width $d_{\rm FFN}$. 
The encoded token for each jet is passed to the per-jet 5-class classifier $C_1$, a two-layer MLP with hidden width $N_{\rm multi}\times d$,
\begin{equation}
    \ell_j^{(5)} =
    C_1\!\left([T_1(z_1,\ldots,z_J)]_j\right)
    \in \mathbb{R}^{5},
    \qquad j=1,\ldots,J .
\end{equation}
After ST1 outputs the five-dimensional logit vector $\ell_j^{(5)}$ for each jet in an event, we rank the valid jets by their logits for the four non-other categories: $b_1$, $W_1$, $b_2$ and $W_2$. For each category, we retain the top four jets with the largest logits. From these candidates, we further select the optimal non-repeating combination that contains exactly (1 $b_1$, 2 $W_1$s, 1 $b_2$, 2 $W_2$s) with the highest ST1 assignment score. This combination is taken as the inferred ST1 assignment for the corresponding event.

\subsubsection{Step-2 Classifier}\label{sec:step2Net}
Following the final classification result of ST1 for each event, the combination (1 $b_1$, 2 $W_1$s, 1 $b_2$, 2 $W_2$s) is orderly organized as
a $N_{\rm token}=6$ input vector, which is
\begin{equation}
    (b_1,q_{1a},q_{1b},b_2,q_{2a},q_{2b}),
\end{equation}
and then it is passed to the ST2 classifier to produce assignment logits $s_{\boldsymbol{\delta}}$ and the corresponding posterior $q_{\boldsymbol{\delta}}$ over the $N_{\rm assign}=4$ down/up assignment hypotheses $\boldsymbol{\delta}=(\delta_{11},\delta_{12},\delta_{21},\delta_{22}), \delta_{ij}\in\{0,1\}$ with $\delta_{11}\neq \delta_{12}$ and $\delta_{21}\neq \delta_{22}$.
The two representations are related by $(\delta_{i1},\delta_{i2})=(Z_i,1-Z_i)$, so the four allowed values of $\boldsymbol{\delta}$ correspond to the four two-bit labels $\mathbf{Z}$.
The network has three components: (i) a shared jet+constituent encoder identical in architecture to ST1 with independently trained parameters, producing fused tokens $z_j\in\mathbb{R}^d$; (ii) an event encoder that builds two branch summaries $r_1,r_2$ and an event context $c$ from the CLS output of a transformer over these summaries; (iii) an ST2 scorer that produces assignment logits $s_{\boldsymbol{\delta}}$ from the tokens and $c$.
An auxiliary conditional diffusion denoiser is trained jointly with the ST2 scorer and shares the same event context.

The six input jets are passed through the shared encoder, yielding the fused tokens $(z_{b_1},z_{q_{1a}},z_{q_{1b}},z_{b_2},z_{q_{2a}},z_{q_{2b}})$.
The event encoder first builds one summary token for each top branch,
\begin{equation}
\begin{split}
    r_1 &= B\!\left([z_{b_1},
        (z_{q_{1a}}+z_{q_{1b}})/2,
        |z_{q_{1a}}-z_{q_{1b}}|]\right),\\
    r_2 &= B\!\left([z_{b_2},
        (z_{q_{2a}}+z_{q_{2b}})/2,
        |z_{q_{2a}}-z_{q_{2b}}|]\right),
\end{split}
\end{equation}
where $B$ is a two-layer ReLU MLP mapping $3d\to d$. A learnable Classification Token (CLS) and the two branch tokens are then processed by a two-layer transformer encoder. The event context $c$ is the CLS output after LayerNorm,
\begin{equation}
    c = \mathrm{LayerNorm}\!\left([T_{\mathrm{event}}([\mathrm{CLS},r_1,r_2])]_{\mathrm{CLS}}\right).
\end{equation}
The ST2 scorer explicitly enumerates the four assignment hypotheses.  For a hypothesis $\boldsymbol{\delta}$, the token triplet used for branch $r$ is
\begin{equation}
(\tilde z_{b_r},\tilde z_{d_r},\tilde z_{u_r}) =
\begin{cases}
(z_{b_r},z_{q_{ra}},z_{q_{rb}}), & \delta_{r1}=0 ~{\rm and }~\delta_{r2}=1,\\
(z_{b_r},z_{q_{rb}},z_{q_{ra}}), & \delta_{r1}=1 ~{\rm and }~\delta_{r2}=0.
\end{cases}
\end{equation}
The branch scoring network then uses this ordered triplet and its pairwise differences,
\begin{equation}
    e_r =
    A_{\mathrm{br}}\!\left([\tilde z_{b_r},\tilde z_{d_r},\tilde z_{u_r},
    \tilde z_{d_r}-\tilde z_{u_r},|\tilde z_{d_r}-\tilde z_{u_r}|,
    \tilde z_{d_r}-\tilde z_{b_r},\tilde z_{u_r}-\tilde z_{b_r}]\right),
\end{equation}
where the signed differences retain the hypothesized down/up ordering, while $|\tilde z_{d_r}-\tilde z_{u_r}|$ is insensitive to the order of the two $W$-daughter tokens. Here $A_{\mathrm{br}}$ is a ReLU MLP mapping $7d\to h_{\mathrm{assign}}\to d$. The ST2 assignment logit and posterior for hypothesis $\boldsymbol{\delta}$ are
\begin{equation}
    s_{\boldsymbol{\delta}} =
    A_{\mathrm{evt}}\!\left([e_1+e_2, |e_1-e_2|, c]\right),
    ~
    q_{\boldsymbol{\delta}} = \frac{\exp s_{\boldsymbol{\delta}}}{\sum_{\boldsymbol{\delta}'}\exp s_{\boldsymbol{\delta}'}} .
\end{equation}
where $A_{\mathrm{evt}}$ is a 2-layer ReLU MLP, mapping $3d\to h_{\mathrm{assign}}\to h_{\mathrm{assign}}/2\to 1$. The predicted assignment is $\arg\max_{\boldsymbol{\delta}} q_{\boldsymbol{\delta}}$.

Alongside the branch assignment module, we incorporate a diffusion network to stabilize the training of ST2. This diffusion network comprises a noiser and a denoiser. The denoiser acts as an auxiliary training branch and shares the event context $c$ across the entire module.
Given an assignment hypothesis $\boldsymbol{\delta}$, we first construct an ordered token sequence:
\begin{equation}
    X_0(\boldsymbol{\delta})=(\tilde z_{b_1},\tilde z_{d_1},\tilde z_{u_1},\tilde z_{b_2},\tilde z_{d_2},\tilde z_{u_2})\in\mathbb{R}^{6\times d}.
\end{equation}
During training, a diffusion timestep $t\in\{0,\ldots,T-1\}$ is drawn by the importance sampler described below, and a noiser with Gaussian distribution $\epsilon\sim\mathcal{N}(0,I)$ is added according to a linear beta schedule,
\begin{equation}
    \beta_t
    =
    \beta_{\min}
    +
    \frac{t}{T-1}
    \left(\beta_{\max}-\beta_{\min}\right),
    \qquad
    \alpha_t=1-\beta_t,
    \qquad
    \bar\alpha_t=\prod_{\tau=0}^{t}\alpha_\tau ,
\end{equation}
\begin{equation}
    X_t(\boldsymbol{\delta})
    =
    \sqrt{\bar\alpha_t}\,X_0(\boldsymbol{\delta})
    +
    \sqrt{1-\bar\alpha_t}\,\epsilon .
\end{equation}
The denoiser predicts the injected noise conditioned on the event context and the assignment hypothesis,
\begin{equation} \hat\epsilon=\epsilon_\theta\!\left(X_t(\boldsymbol{\delta}),t,c,\boldsymbol{\delta}\right) \in\mathbb{R}^{6\times d}.
\end{equation}
Here, the denoiser is implemented as a small transformer encoder, which is built as
\begin{equation}
\epsilon_\theta\!\left(X_t(\boldsymbol{\delta}),t,c,\boldsymbol{\delta}\right)
    =
    W_{\mathrm{out}}\,
    T_{\mathrm{diff}}\!\left[
    X_t(\boldsymbol{\delta})
    + e_t
    + e_{\mathrm{role}}
    + W_c c
    + e_{\boldsymbol{\delta}}
    \right],
\end{equation}
where the timestep embedding $e_t$ is obtained by first applying a fixed sinusoidal encoding $\gamma(t)\in\mathbb{R}^d$, followed by a two-layer MLP with hidden width $d$ and SiLU activation, which is $e_t = E_t(\gamma(t))\in\mathbb{R}^d$, and then broadcast to all six tokens; $e_{\mathrm{role}}\in\mathbb{R}^{6\times d}$ is formed by repeating three learnable role embeddings $e_b,e_d,e_u\in\mathbb{R}^d$ across the two branches, following the ordered role pattern $(b,d,u;\,b,d,u)$.  The projected event-context embedding $W_c c$ and the assignment-hypothesis embedding $e_{\boldsymbol{\delta}}$ are also broadcast to all six tokens.
The output layer is a token-wise linear projection $W_{\mathrm{out}}\in\mathbb{R}^{d\times d}$, applied without activation, yielding a predicted noise tensor $\hat\epsilon\in\mathbb{R}^{6\times d}$.

The diffusion assignment score $\mathcal{G}_{\boldsymbol{\delta}}$ is the negative denoising error for hypothesis $\boldsymbol{\delta}$, averaged over $K_{\rm diff}$ Monte-Carlo draws,
\begin{equation}\label{eq:G-delta}
    \mathcal{G}_{\boldsymbol{\delta}}
    =
    -\frac{1}{K_{\rm diff}}\sum_{k=1}^{K_{\rm diff}}
    \left(\left\|
    \epsilon^{(k)}-\epsilon_\theta\!\left(X_{t^{(k)}}(\boldsymbol{\delta}),t^{(k)},c,\boldsymbol{\delta}\right)
    \right\|_2\right)^2,
\end{equation}
where $\left\| \cdot \right\|_2$ is the matrix Frobenius norm. In Eq.~\eqref{eq:G-delta}, $t^{(k)}$ is sampled from the adaptive distribution $p(t)$ during training and uniformly at inference.

For a training event $e$, let $\mathbf{Z}_e$ denote its true two-bit down/up assignment. Since the ST2 loss below compares $\mathcal{G}_{e,\mathbf{Z}_e}$ with the competing $\mathcal{G}_{e,\boldsymbol{\delta}}$, the same pair $(t^{(k)},\epsilon^{(k)})$ is used for all four hypotheses in each Monte Carlo draw, reducing common-mode fluctuations in the score differences~\cite{Li_2023_ICCV}. After a uniform-sampling warm-up, $p(t)$ is updated from recent raw denoising errors and assigns larger probabilities to timesteps with larger errors. The diagnostic denoising-loss estimate uses the inverse-probability factor $1/[T p(t)]$~\cite{pmlr-v139-nichol21a}, while $\mathcal{G}_{e,\boldsymbol{\delta}}$ and the margin loss use the sampled raw denoising errors. We use $K_{\rm diff}=4$ draws during both training and inference.

\subsection{Loss Functions of the Network}
For ST1 training, let $y_{ej}^{(5)}$ denote the truth five-class label of valid jet $j$ in event $e$. The loss is the cross-entropy (CE) averaged over valid jets inside each event. The labels have a branch-exchange ambiguity: swapping the two top quark maps
\begin{equation}
    b_1\leftrightarrow b_2,\qquad W_1\leftrightarrow W_2,
    \qquad \mathrm{other}\mapsto \mathrm{other}.
\end{equation}
which enforces a permutation invariance for the ST1 loss function. This loss function is formulated as
\begin{equation}
    \mathcal{L}_{\rm ST1}=\frac{1}{N}\sum_{e=1}^{N}
    \min\left[
        \frac{1}{n_e}\sum_{j\in\mathcal{V}_e}
        \mathrm{CE}\!\left(\ell_{ej}^{(5)},y_{ej}^{(5)}\right),
        \frac{1}{n_e}\sum_{j\in\mathcal{V}_e}
        \mathrm{CE}\!\left(\ell_{ej}^{(5)},\pi\!\left(y_{ej}^{(5)}\right)\right)
    \right],
\end{equation}
where $N$ is the total training events, $\mathcal{V}_e$ is the valid-jet set, $n_e=|\mathcal{V}_e|$, and $\pi$ is the branch-swap map.

For ST2 training, the classifier loss is the ordinary four-class cross-entropy,
\begin{equation}
    \mathcal{L}_{\mathrm{CE},2}
    =
    -\frac{1}{N}\sum_{e=1}^{N}\log q_{e,\mathbf{Z}_e}.
\end{equation}
The total ST2 loss combines the classifier cross-entropy with a diffusion margin term, which is
\begin{equation}\label{eq:step2loss}
    \mathcal{L}_{\rm ST2}
    =
    \mathcal{L}_{\mathrm{CE},2}
    +\frac{\lambda_{\mathrm{margin}}}{N}
    \sum_{e=1}^{N}
    \left[
        m_{\rm bias}
        -\mathcal{G}_{e,\mathbf{Z}_e}
        +\max_{\boldsymbol{\delta}\ne\mathbf{Z}_e}
        \mathcal{G}_{e,\boldsymbol{\delta}}
    \right]_+ .
\end{equation}
Here $\lambda_{\rm margin}$ is the weight of the diffusion margin term, $\mathcal{G}_{e,\boldsymbol{\delta}}$ denotes the diffusion assignment score of hypothesis $\boldsymbol{\delta}$ in event $e$, as defined in Eq.~\eqref{eq:G-delta}, and $[\cdot]_+=\max(\cdot,0)$. The plain denoising mean-squared error is not back-propagated. The diffusion branch therefore enters training only through the margin term, which requires the true-assignment score $\mathcal{G}_{e,\mathbf{Z}_e}$ to exceed the largest competing score by $m_{\rm bias}$. The diffusion branch thus supplies an auxiliary ranking signal to the shared event representation and provides additional diffusion cut variables, with the final down-type assignment obtained from the ST2 scorer.

\subsection{Assignment Scores and Cut Variables}\label{sec:assignmentScore}
This subsection defines the event-level assignment scores used for inference and the nine ML cut variables used for downstream cuts: four from ST1, three from ST2, and two from the diffusion assignment scores $\mathcal{G}_{\boldsymbol{\delta}}$.

Let $Y\in\mathcal{A}_{\rm legal}$ denote a legal ST1 six-jet assignment, and let $\boldsymbol{\delta}=(\delta_{11},\delta_{12},\delta_{21},\delta_{22})$ denote a ST2 down-type quark assignment hypothesis.
The inferred ST1 and ST2 assignments are
\begin{equation}
    \hat Y = \arg\max_{Y\in\mathcal{A}_{\rm legal}} S_1(Y;X),
    \qquad
    \hat{\boldsymbol{\delta}} = \arg\max_{\boldsymbol{\delta}} S_2(\boldsymbol{\delta};X),
\end{equation}
where $S_1$ and $S_2$ are the {\it assignment scores} for each event defined below.

For ST1, a legal assignment $Y$ selects two $b$-jets and four $W$-daughter jets from the valid-jet set $\mathcal{V}$.
To keep the enumeration tractable, each role is restricted to the $k_{\rm top}$ jets with the highest corresponding logit.
Throughout this subsection, $\mathcal{A}_{\rm legal}$ denotes the restricted legal candidate set obtained after the $k_{\rm top}$ truncation.
For a given assignment $Y$, let $y_j(Y)$ denote the induced five-class label of each valid jet $j$.
The ST1 assignment score combines the logit term and a mass prior term,
\begin{equation}
    S_1(Y;X)=S_{\mathrm{ML}}(Y;X)+S_{\mathrm{mass}}(Y;X),
\end{equation}
with
\begin{equation}
    S_{\mathrm{ML}}(Y;X)
    =\sum_{j\in\mathcal{V}}\log\left\{\left[\mathrm{softmax}\!\left(\ell_j^{(5)}\right)\right]_{y_j(Y)}\right\},
\end{equation}
\begin{equation}
    S_{\mathrm{mass}}(Y;X)
    =-\lambda_W^{\rm mass}\sum_{r=1}^{2}\left(\frac{m_{W,r}-m_W}{\sigma_W^{\rm prior}}\right)^{\!2}
    -\lambda_t^{\rm mass}\sum_{r=1}^{2}\left(\frac{m_{t,r}-m_t}{\sigma_t^{\rm prior}}\right)^{\!2}.
\end{equation}
The ST1 assignment posterior is defined by the softmax of $S_1$,
\begin{equation}
    Q_1(Y|X)
    =\frac{\exp S_1(Y;X)}{\sum_{Y'\in\mathcal{A}_{\rm legal}}\exp S_1(Y';X)} .
\end{equation}
The four ST1 cut variables are
\begin{equation}
\begin{aligned}
    Q_{\rm ST1}^{\rm avg}&=\frac{1}{|\mathcal{V}|}\log\sum_{Y\in\mathcal{A}_{\rm legal}}\exp S_1(Y;X),\\
    Q_{\rm ST1}^{\rm max}&=\max_{Y\in\mathcal{A}_{\rm legal}} Q_1(Y|X),\\
    Q_{\rm ST1}^{\rm etp}&=-\sum_{Y\in\mathcal{A}_{\rm legal}} Q_1(Y|X)\log Q_1(Y|X),\\
    Q_{\rm ST1}^{\rm gap}&=\bigl[S_1(Y_{(1)};X)-S_1(Y_{(2)};X)\bigr]/|\mathcal{V}|,
\end{aligned}
\end{equation}
where $Y_{(1)},Y_{(2)}$ are the assignments with the largest and second-largest $S_1$.

For ST2, the assignment score is the assignment logit defined in Sec.~\ref{sec:step2Net},
\begin{equation}
    S_2(\boldsymbol{\delta};X)\equiv s_{\boldsymbol{\delta}},
    \qquad
    Q_2(\boldsymbol{\delta}|X)
    \equiv q_{\boldsymbol{\delta}}
    =\frac{\exp S_2(\boldsymbol{\delta};X)}{\sum_{\boldsymbol{\delta}'}\exp S_2(\boldsymbol{\delta}';X)} .
\end{equation}
The three ST2 cut variables are
\begin{equation}
\begin{aligned}
    Q_{\rm ST2}^{\rm max}&=\max_{\boldsymbol{\delta}} Q_2(\boldsymbol{\delta}|X),\\
    Q_{\rm ST2}^{\rm etp}&=-\sum_{\boldsymbol{\delta}} Q_2(\boldsymbol{\delta}|X)\log Q_2(\boldsymbol{\delta}|X),\\
    Q_{\rm ST2}^{\rm gap}&=S_{2,(1)}-S_{2,(2)},
\end{aligned}
\end{equation}
with $S_{2,(1)}\ge S_{2,(2)}$ the largest two ST2 assignment scores.

The two diffusion cut variables are computed from the diffusion assignment scores $\mathcal{G}_{\boldsymbol{\delta}}$:
\begin{equation}
    Q_{\rm diff}^{\rm max}=\max_{\boldsymbol{\delta}} \mathcal{G}_{\boldsymbol{\delta}},
    \qquad
    Q_{\rm diff}^{\rm gap}=\mathcal{G}_{(1)}-\mathcal{G}_{(2)} ,
\end{equation}
with $\mathcal{G}_{(1)}\ge \mathcal{G}_{(2)}$ the largest two diffusion assignment scores.

\begin{table*}[t]
\centering
\caption{Hyperparameters of the two-stage reconstruction network, training, and inference. The total trainable parameter counts of ST1 and ST2 are $7.08\times10^6$ and $7.81\times10^6$, respectively.}
\label{tab:hyperparameters}
\small
\renewcommand{\arraystretch}{1.05}
\begin{tabular}{llcl}
\hline
Category & Symbol & Value & Description \\
\hline

\multirow{6}{*}{Input \& padding}
& $d_{\rm jet}$ & 14 & jet feature dimension \\
& $d_{\rm const}$ & 10 & constituent feature dimension \\
& $N_{\rm const}^{\max}$ & 40 & max constituents per jet \\
& $J$ & 17 & max jets per event \\
& $N_{\rm token}$ & 6 & selected jets (step 2) \\
& $N_{\rm assign}$ & 4 & down/up assignment classes \\
\hline

\multirow{6}{*}{\shortstack[l]{Shared encoder\\and fusion}}
& $d$ & 320 & token width \\
& $h_{\rm w}$ & 2 & MLP hidden-width multiplier \\
& $K_{\rm GNN}$ & 8 & EdgeConv nearest neighbours \\
& $c_1$ & 32 & first EdgeConv block width \\
& $c_2$ & 64 & second EdgeConv block width \\
& $g_0$ & $-2$ & GNN gate initial value \\
\hline

\multirow{4}{*}{Step-1 classifier}
& $N_{\rm layer}$ & 6 & transformer layers \\
& $N_{\rm head}$ & 10 & attention heads \\
& $d_{\rm FFN}$ & 1024 & feed-forward width \\
& $N_{\rm multi}$ & 2 & classifier hidden-width multiplier \\
\hline

\multirow{4}{*}{\shortstack[l]{Step-2 event\\encoder \& scorer}}
& $N_{\rm layer}^{\rm event}$ & 2 & event transformer layers \\
& $N_{\rm head}^{\rm event}$ & 10 & event transformer heads \\
& $d_{\rm ff}^{\rm event}$ & 1024 & event transformer FF width \\
& $h_{\rm assign}$ & 320 & assignment scorer hidden width \\
\hline

\multirow{7}{*}{\shortstack[l]{Conditional\\diffusion}}
& $T$ & 50 & diffusion timesteps \\
& $\beta_{\min}$ & $10^{-4}$ & initial beta \\
& $\beta_{\max}$ & $2\times 10^{-2}$ & final beta \\
& $N_{\rm layer}^{\rm diff}$ & 3 & denoiser transformer layers \\
& $N_{\rm head}^{\rm diff}$ & 10 & denoiser transformer heads \\
& $d_{\rm ff}^{\rm diff}$ & 1024 & denoiser FF width \\
& $K_{\rm diff}$ & 4 & paired $(t,\epsilon)$ draws averaged per score \\
\hline

\multirow{2}{*}{Loss weights}
& $\lambda_{\rm margin}$ & 50 & margin loss weight \\
& $m_{\rm bias}$ & 0.02 & margin threshold \\
\hline

\multirow{2}{*}{Dropout}
& $p_{\rm drop}^{(1)}$ & 0.10 & step-1 transformer and classifier dropout \\
& $p_{\rm drop}^{(2)}$ & 0.15 & ST2 encoder, scorer, and denoiser dropout \\
\hline

\multirow{3}{*}{\shortstack[l]{$B_{\rm ATL}$\\$\chi^2$ score}}
& $m_W$ & 80.4\,GeV & $W$ mass reference \\
& $\sigma_t$ & 17.6\,GeV & top mass resolution \\
& $\sigma_W$ & 9.3\,GeV & $W$ mass resolution \\
\hline

\multirow{6}{*}{\shortstack[l]{Mass prior}}
& $k_{\rm top}$ & 4 & candidates per class in enumeration \\
& $m_t$ & 172.5\,GeV & top mass reference \\
& $\lambda_W^{\rm mass}$ & 0.15 & $W$ mass prior weight \\
& $\lambda_t^{\rm mass}$ & 0.10 & top mass prior weight \\
& $\sigma_W^{\rm prior}$ & 45.0\,GeV & $W$ mass prior width \\
& $\sigma_t^{\rm prior}$ & 85.0\,GeV & top mass prior width \\
\hline
\end{tabular}
\end{table*}

\subsection{Training}
The matched dataset is split into training, validation, and test sets in the ratio $0.8/0.1/0.1$ with a fixed random seed.
The two stages are trained independently; ST2 uses the same split indices as ST1.
The stages are trained independently rather than end-to-end because the down/up labels are substantially more ambiguous than the coarse ST1 labels. Keeping the ST2 loss separate prevents this ambiguous objective from degrading the first-stage six-jet assignment training.

Both stages are optimized with AdamW~\cite{loshchilov2018decoupled} using an initial learning rate of $10^{-4}$ and weight decay $10^{-4}$, with batch size 32.
The learning rate is reduced by a factor of $0.5$ whenever the validation loss plateaus for 3 consecutive epochs.
ST1 is trained for 100 epochs.
ST2 is trained for 40 epochs.
In both cases, the checkpoint with the lowest validation loss is selected for inference.

ST2 is trained on truth-matched jets.  For each event, the two $W$-daughter jets in each branch are randomly ordered as $(q_{ia},q_{ib})$, and the target label specifies which one is the down-type jet.
At inference time, however, the six input jets are instead taken from the highest-scoring legal assignment found by ST1.

\section{Results and Discussions}
\label{sec:results}
We now evaluate the two-stage reconstruction from three complementary perspectives. We first examine how the diffusion training objective affects the down assignment score, then study the background rejection and reconstruction quality obtained with the ML cut variables, and finally quantify how much spin-correlation information is retained by the reconstructed down-type jet directions.

\subsection{Margin Loss vs Diffusion Loss}

The conditional diffusion head can in principle rank the four assignment hypotheses through the diffusion assignment score $\mathcal{G}_{\boldsymbol{\delta}}$ (Sec.~\ref{sec:step2Net}), with $\hat{\boldsymbol{\delta}}=\arg\max_{\boldsymbol{\delta}} \mathcal{G}_{\boldsymbol{\delta}}$, in the spirit of diffusion classifier constructions~\cite{Li_2023_ICCV}.  The performance of this diffusion ranking depends largely on the loss function.

Without the margin objective, all tested diffusion variants stay close to the random baseline of $0.25$ for four hypotheses (median $0.267$, best $0.280$).  We tested the plain denoising diffusion objective~\cite{NEURIPS2020_4c5bcfec} together with several common modifications: averaging $\mathcal{G}_{\boldsymbol{\delta}}$ over multiple noise draws~\cite{Li_2023_ICCV}, sampling timesteps more often when their recent denoising errors are larger~\cite{pmlr-v139-nichol21a}, predicting the alternative $v$ target instead of the noise~\cite{salimans2022progressive}, rescaling the diffusion network input and output as in the EDM framework~\cite{karras2022elucidating}, randomly replacing the assignment condition by a null condition during training~\cite{ho2021classifierfree}, using the cosine noise schedule~\cite{pmlr-v139-nichol21a}, and reweighting the loss by the signal to noise ratio~\cite{Hang_2023_ICCV}.  None of these made $\mathcal{G}_{\boldsymbol{\delta}}$ useful for ranking the four assignment hypotheses.

\begin{figure*}
    \centering
    \includegraphics[width=0.32\linewidth]{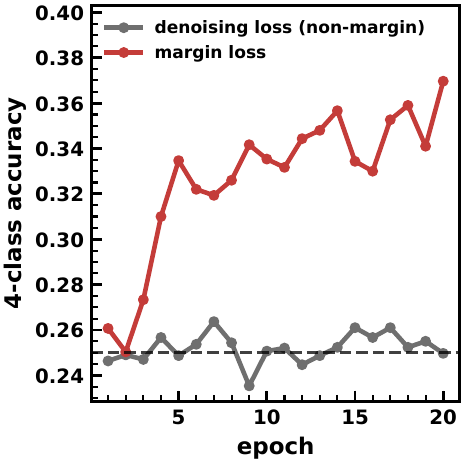}\hfill
    \includegraphics[width=0.32\linewidth]{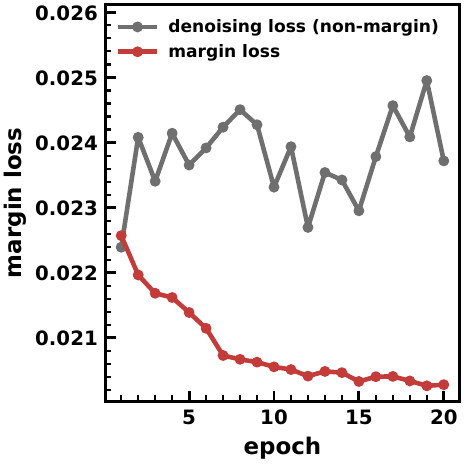}\hfill
    \includegraphics[width=0.32\linewidth]{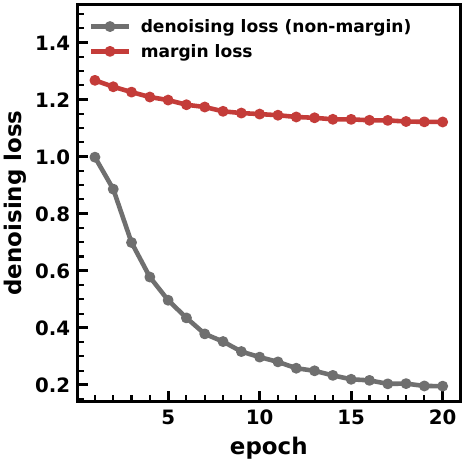}
    \caption{Comparison of the two diffusion head training objectives versus training epoch, at otherwise identical settings: the plain denoising loss (gray) and the margin loss (red).  \textbf{Left:} validation four class accuracy of $\arg\max_{\boldsymbol{\delta}} \mathcal{G}_{\boldsymbol{\delta}}$; the denoising objective stays at the random level (dashed, $0.25$) while the margin objective becomes discriminative.  \textbf{Middle:} the margin loss term, only the margin objective drives it down.  \textbf{Right:} the plain objective reduces the denoising loss without improving assignment accuracy, whereas the margin objective yields discriminative $\mathcal{G}_{\boldsymbol{\delta}}$ values despite a larger denoising loss.  This scan motivates training the deployed diffusion head through the margin objective alone.}
    \label{fig:margin-loss}
\end{figure*}

The decisive ingredient is the margin term of Eq.~(\ref{eq:step2loss}), which directly optimizes the ordering of the true-assignment diffusion score $\mathcal{G}_{e,\mathbf{Z}_e}$ above the competing scores $\mathcal{G}_{e,\boldsymbol{\delta}}$.  With an otherwise identical network, the margin objective lifts the accuracy of $\arg\max_{\boldsymbol{\delta}}\mathcal{G}_{\boldsymbol{\delta}}$ well above random, whereas training the same network only to predict the injected diffusion noise leaves the accuracy at the random level (Fig.~\ref{fig:margin-loss}).  Moreover, the two objectives are decoupled: training only to predict the injected diffusion noise steadily lowers the denoising loss yet never moves the assignment accuracy away from random, whereas the margin objective retains a comparatively high denoising loss but produces a $\mathcal{G}_{\boldsymbol{\delta}}$ that discriminates assignment hypotheses.  Optimizing denoising accuracy is thus orthogonal to the assignment task.

Increasing the number of paired $(t,\epsilon)$ samples used to estimate $\mathcal{G}_{\boldsymbol{\delta}}$ from $K_{\rm diff}=4$ to $K_{\rm diff}=256$ does not improve assignment accuracy, indicating that the remaining limitation is not Monte Carlo noise in $\mathcal{G}_{\boldsymbol{\delta}}$.  We therefore use $K_{\rm diff}=4$ and train the deployed ST2 diffusion head only through the margin term, without back-propagating the denoising mean-squared error.  This scan supports the auxiliary head design of the deployed network.

\subsection{Cut Variables and Reconstruction Quality}
The $B_{\rm ATL}$ baseline is defined in Sec.~\ref{sec:background}, and the ML cut variables entering $\mathcal{V}_{\rm ML}$ are defined in Sec.~\ref{sec:assignmentScore}. Let $\epsilon^{B_{\rm ATL}}_a$ denote the retention fraction of sample $a\in\{\rm sig,\,unm,\, QCD\}$ corresponding to \{signal, unmatched, QCD\} respectively after the $B_{\rm ATL}$ baseline cut alone, before imposing the validity of the ML reconstruction.  The quantity $\epsilon^{B_{\rm ATL}+\mathcal{V}_{\rm ML}}_a$ denotes the retention fraction after additionally requiring a valid two-step ML inference result and passing the ML selection $\mathcal{V}_{\rm ML}$, where $\mathcal{V}_{\rm ML}$ is any combination of the ML cut variables used in the performance scan; even the limiting case $\mathcal{V}_{\rm ML}=0$ therefore includes the inference-validity requirement and is not identical to $\epsilon^{B_{\rm ATL}}_a$.  With the preselection ratio $R_{\rm unm/sig}^{\rm pre}$ in Eq. \eqref{eq:Rpre}, the expected background-to-signal ratio after the joint cut decomposes as
\begin{equation}\label{eq:bkg-sig-ratio}
    R_{\rm bkg/sig}^{B_{\rm ATL}+\mathcal{V}_{\rm ML}}=\frac{N_{\rm bkg}^{B_{\rm ATL}+\mathcal{V}_{\rm ML}}}{N_{\rm sig}^{B_{\rm ATL}+\mathcal{V}_{\rm ML}}}
    =\frac{N_{\rm unm}^{B_{\rm ATL}+\mathcal{V}_{\rm ML}}+N_{\rm QCD}^{B_{\rm ATL}+\mathcal{V}_{\rm ML}}}{N_{\rm sig}^{B_{\rm ATL}+\mathcal{V}_{\rm ML}}},
\end{equation}
with
\begin{equation}\label{eq:bkg-sig-components}
\begin{aligned}
    &R_{\rm unm/sig}^{B_{\rm ATL}+\mathcal{V}_{\rm ML}}=\frac{N_{\rm unm}^{B_{\rm ATL}+\mathcal{V}_{\rm ML}}}{N_{\rm sig}^{B_{\rm ATL}+\mathcal{V}_{\rm ML}}}
    =R_{\rm unm/sig}^{\rm pre}\,
    \frac{\epsilon^{B_{\rm ATL}+\mathcal{V}_{\rm ML}}_{\rm unm}}{\epsilon^{B_{\rm ATL}+\mathcal{V}_{\rm ML}}_{\rm sig}} ,\\
&R_{\rm QCD/sig}^{B_{\rm ATL}+\mathcal{V}_{\rm ML}}=\frac{N_{\rm QCD}^{B_{\rm ATL}+\mathcal{V}_{\rm ML}}}{N_{\rm sig}^{B_{\rm ATL}+\mathcal{V}_{\rm ML}}}
    =
    \frac{R^{B_{\rm ATL}}_{{\rm QCD}/t\bar t}\,
    \bigl(R_{\rm unm/sig}^{\rm pre}\,\epsilon^{B_{\rm ATL}}_{\rm unm}+\epsilon^{B_{\rm ATL}}_{\rm sig}\bigr)}
    {\epsilon^{B_{\rm ATL}}_{\rm QCD}}\,
    \frac{\epsilon^{B_{\rm ATL}+\mathcal{V}_{\rm ML}}_{\rm QCD}}{\epsilon^{B_{\rm ATL}+\mathcal{V}_{\rm ML}}_{\rm sig}}.
\end{aligned}
\end{equation}
The corresponding signal sample purity is $P_{\rm pur}=1/(1+R_{\rm bkg/sig}^{B_{\rm ATL}+\mathcal{V}_{\rm ML}})$.

We evaluate the cut performance using the test split of the fully matched signal sample, together with the full unmatched $t\bar t$ and QCD samples used in the inference cache.  The signal test split is used only to estimate the matched-signal efficiency and reconstruction observables; the normalization of the signal, unmatched, and QCD components follows the equations above.  The baseline selection is always $B_{\rm ATL}$, and the additional ML selection is denoted by $\mathcal{V}_{\rm ML}$.  The numerical impact of the pure baseline, the valid-inference requirement, and the representative ML working point is summarized in Table~\ref{tab:working-points} after the spin-analysis observables are defined below.

\begin{figure}
    \centering
    \includegraphics[width=0.48\linewidth]{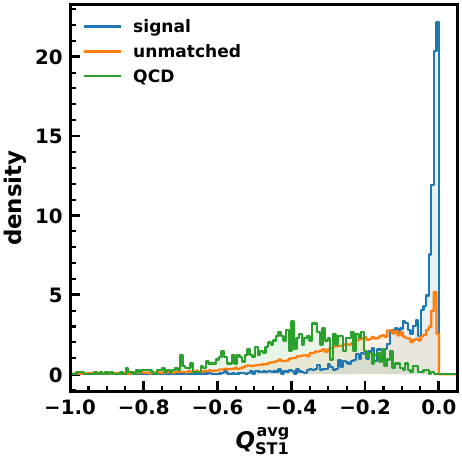}
    \caption{Distribution of the step-1 average log evidence, \texttt{step1\_avg}, for signal, unmatched $t\bar t$, and QCD events after the $B_{\rm ATL}$ baseline and the valid-inference requirement.  This variable is used as the representative one-dimensional cut in the benchmark working point below.}
    \label{fig:step1-avg-distribution}
\end{figure}

As a representative ML cut variable, Fig.~\ref{fig:step1-avg-distribution} shows the distribution of $Q_{\rm ST1}^{\rm avg}$ after the $B_{\rm ATL}$ baseline and the valid-inference requirement.  The signal distribution is shifted toward larger values, while the unmatched $t\bar t$ and QCD samples populate lower-evidence regions more strongly.

\begin{figure}
    \centering
    \includegraphics[width=0.47\linewidth]{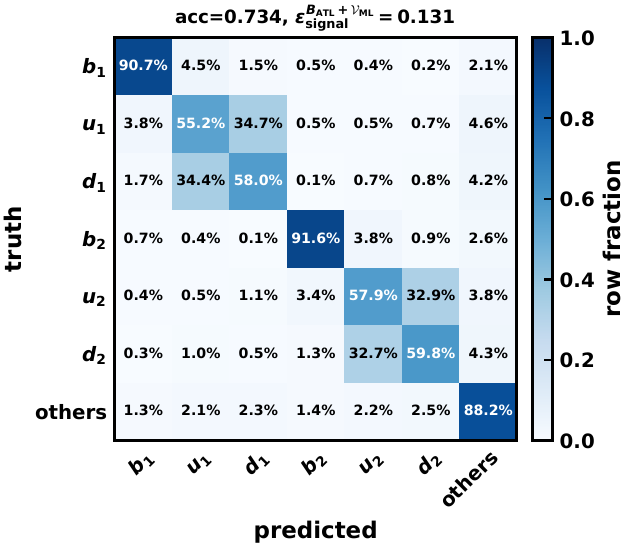}\hfill
    \includegraphics[width=0.47\linewidth]{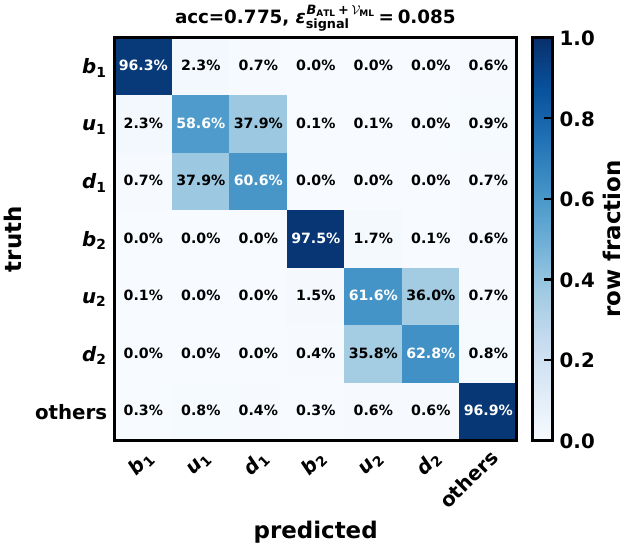}
    \caption{Seven-class jet-label confusion matrices for the two-step reconstruction on the signal test split. \textbf{Left:} $B_{\rm ATL}$ plus valid inference.  \textbf{Right:} the same baseline with an additional representative $Q_{\rm ST1}^{\rm avg}\ge -0.0889$ cut, giving $\epsilon_{\rm signal}^{B_{\rm ATL}+\mathcal{V}_{\rm ML}}\simeq0.0845$.  Rows are truth labels and columns are predicted labels.}
    \label{fig:confusion-matrices}
\end{figure}

Figure~\ref{fig:confusion-matrices} shows the corresponding seven-class jet-label confusion matrices on the signal test split.  The entries are row-normalized over the truth labels $(b_1,u_1,d_1,b_2,u_2,d_2,\mathrm{others})$, with the same overall branch-interchange ambiguity allowed as in the final six-parton reconstruction.  The representative $Q_{\rm ST1}^{\rm avg}\ge -0.0889$ cut corresponds to $\epsilon_{\rm signal}^{B_{\rm ATL}+\mathcal{V}_{\rm ML}}\simeq0.0845$; it selects a cleaner subset, raising the per-jet label accuracy from $0.734$ to $0.775$, while the remaining dominant confusion is still the same-branch $u/d$ interchange.

\subsection{Spin-analysis Performance}
For each working point we report the retained signal efficiency $\epsilon_{\rm signal}^{B_{\rm ATL}+\mathcal{V}_{\rm ML}}$, the calibrated background-to-signal ratio $R_{\rm bkg/sig}^{B_{\rm ATL}+\mathcal{V}_{\rm ML}}$ defined above, the purity-weighted six-parton reconstruction fraction,
\begin{equation}
    P_{\rm pur}\cdot{\rm Acc}_{6h}
    =
    \frac{1}{1+R_{\rm bkg/sig}^{B_{\rm ATL}+\mathcal{V}_{\rm ML}}}\,{\rm Acc}_{6h},
\end{equation}
and the effective spin-analysis factor $\eta_\kappa$ defined below. Here, ${\rm Acc}_{6h}$ is the fraction of selected signal events for which all six top-decay partons are assigned to the correct jets, allowing an overall interchange of the two top branches.

The down-type quark is a maximal spin analyzer ($\beta_d\simeq 1$), so the measurable $t\bar t$ spin correlation is set by how closely the network-predicted down-jet direction tracks the true down-quark direction. For each selected event $e$ we define the event-level direction-matching score
\begin{equation}
    \kappa^{(e)}
    =
    \max_{\pi\in \mathfrak{S}_2}
    \frac{1}{2}\sum_{r=1,2}
    \hat n^{\rm pred}_{e,r}\cdot \hat n^{\rm truth}_{e,\pi(r)} ,
\end{equation}
where $\hat n^{\rm pred}_{e,r}$ is the predicted down-jet direction and $\hat n^{\rm truth}_{e,r}$ is the generator-level down-parton direction, both evaluated in the truth-top rest frame of branch $r$, with $p_t=p_b+p_u+p_d$. The maximization over $\mathfrak{S}_2$ accounts for the overall interchange of the two top branches, as in ${\rm Acc}_{6h}$.  

We then average this score separately on the signal and the unmatched samples,
\begin{equation}
    \kappa_{\rm sig}=\frac{1}{N_{\rm sig}}\sum_{e~\in~\{\rm sig\}}\kappa^{(e)},
    \qquad
    \kappa_{\rm unm}=\frac{1}{N_{\rm unm}}\sum_{e~\in~\{\rm unm\}}\kappa^{(e)},
\end{equation}
where the sums run over events that pass the joint cut in the last subsection. The unmatched sample lacks a complete jet-parton match but retains generator-level top-decay momenta, so it contributes to the spin-carrying numerator in Eq.~\eqref{eq:etakappa}; QCD contributes only through the denominator.
We combine these effects into the effective spin-analysis factor
\begin{equation}\label{eq:etakappa}
    \eta_\kappa
    =
    \frac{\kappa_{\rm sig}^2+(R_{\rm unm/sig}^{B_{\rm ATL}+\mathcal{V}_{\rm ML}})\,\kappa_{\rm unm}^2}
         {1+R_{\rm bkg/sig}^{B_{\rm ATL}+\mathcal{V}_{\rm ML}}}.
\end{equation}
\begin{table}[t]
    \centering
    \begin{tabular}{lcccc}
        \hline
        Selection & $\epsilon_{\rm signal}$ & $N_{\rm bkg}/N_{\rm sig}$ & $\eta_\kappa$ & $P_{\rm pur}\cdot{\rm Acc}_{6h}$ \\
        \hline
        $B_{\rm ATL}$ & $0.132$ & $7.69$ & -- & -- \\
        $B_{\rm ATL}+0$ & $0.131$ & $7.45$ & $0.0607$ & $0.0348$ \\
        $B_{\rm ATL}+\mathcal{V}_{\rm rep}$ & $0.0845$ & $0.631$ & $0.255$ & $0.218$ \\
        \hline
    \end{tabular}
    \caption{Three reference selections for the reconstruction performance.  The first row applies only the ATLAS-like baseline.  The second row additionally requires a valid two-step ML reconstruction but no extra ML cut.  The third row adds the representative $Q_{\rm ST1}^{\rm avg}\ge -0.0889$ cut, denoted by $\mathcal{V}_{\rm rep}$.}
    \label{tab:working-points}
\end{table}

In Table~\ref{tab:working-points}, the valid-reconstruction requirement changes the signal efficiency from $0.132$ to $0.131$.  The representative $Q_{\rm ST1}^{\rm avg}\ge -0.0889$ cut then suppresses $N_{\rm bkg}/N_{\rm sig}$ and preferentially retains events with higher spin-analysis quality.

\begin{figure}[htpb]
    \centering
    \includegraphics[width=0.95\linewidth]{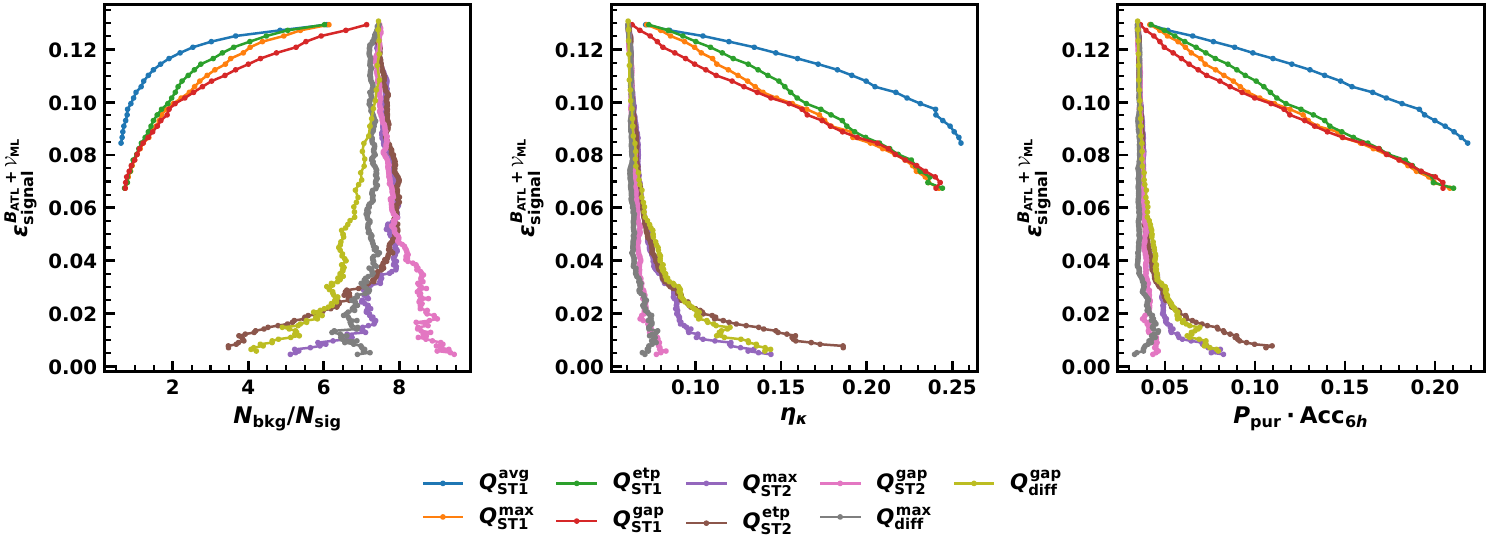}
    \caption{One-dimensional ML cut scans for the nine variables in Sec.~\ref{sec:assignmentScore}.  The three panels show the retained signal efficiency versus $N_{\rm bkg}/N_{\rm sig}$, $\eta_\kappa$, and $P_{\rm pur}\cdot{\rm Acc}_{6h}$, respectively.  Points with fewer than 20 surviving QCD events are omitted.}
    \label{fig:cut-performance}
\end{figure}

Figure~\ref{fig:cut-performance} summarizes the one-dimensional threshold scans using the nine ML cut variables defined in Sec.~\ref{sec:assignmentScore}.  For each variable, the threshold is tightened monotonically and the resulting working points are plotted as the retained signal efficiency $\epsilon_{\rm signal}^{B_{\rm ATL}+\mathcal{V}_{\rm ML}}$ versus the corresponding $N_{\rm bkg}/N_{\rm sig}$, $\eta_\kappa$, or $P_{\rm pur}\cdot{\rm Acc}_{6h}$ value.  Points with fewer than 20 surviving QCD events are omitted to avoid statistically unstable ratios.

\begin{figure}[htbp]
    \centering
    \includegraphics[width=0.95\linewidth]{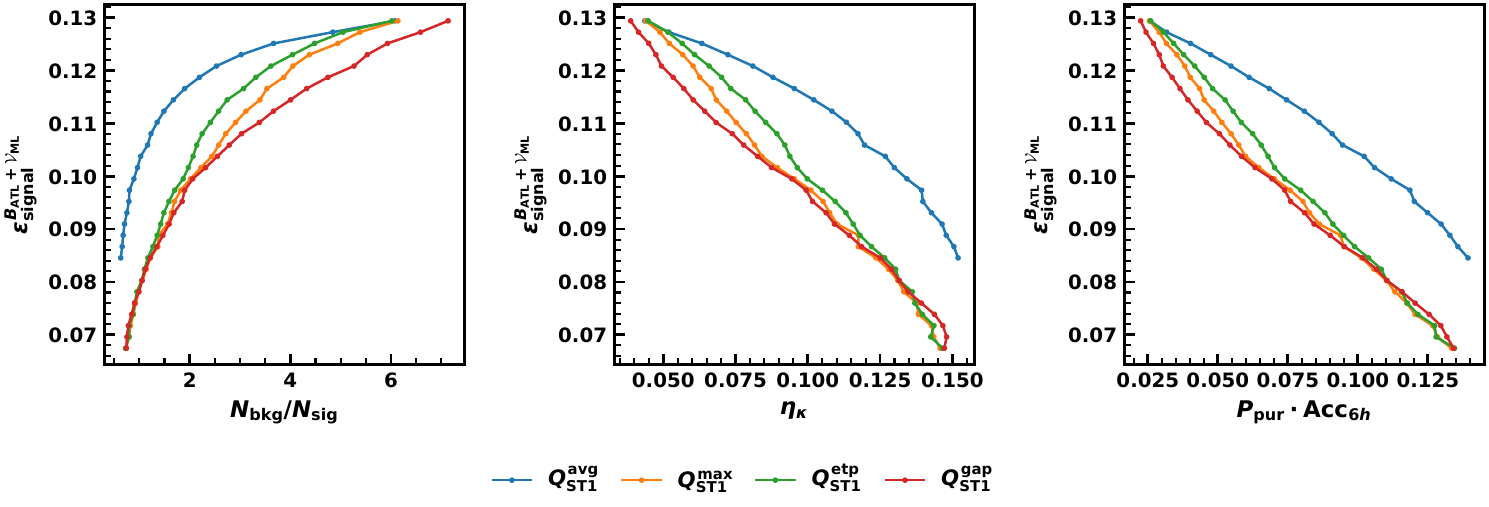}
    \caption{Step 1 only benchmark with random step 2 down/up labels.  The three panels use the same threshold scan format as Fig.~\ref{fig:cut-performance}: $\epsilon_{\rm signal}^{B_{\rm ATL}+\mathcal{V}_{\rm ML}}$ is plotted against $N_{\rm bkg}/N_{\rm sig}$, $\eta_\kappa$, and $P_{\rm pur}\cdot {\rm Acc}_{6h}$.  The benchmark scans only step 1 cut variables and apply the same requirement to the surviving QCD count.  Differences from Fig.~\ref{fig:cut-performance} quantify the contribution of the learned step 2 down/up classifier.}
    \label{fig:cut-performance-compare}
\end{figure}

Figure~\ref{fig:cut-performance-compare} compares the learned ST2 down/up assignment with a ST1 only benchmark.  In this benchmark, the six-jet assignment selected by ST1 is kept fixed, but the ST2 down/up label is replaced by a uniform random choice among the four possibilities.  The benchmark uses the same $N_{\rm bkg}/N_{\rm sig}$, $\epsilon_{\rm signal}^{B_{\rm ATL}+\mathcal{V}_{\rm ML}}$, $\eta_\kappa$, and $P_{\rm pur}\cdot {\rm Acc}_{6h}$ definitions as above, but scans only ST1 cut variables.  For the same ST1 event selection, the calibrated $N_{\rm bkg}/N_{\rm sig}$ curve is therefore unchanged.  The comparison therefore focuses on $\eta_\kappa$ and $P_{\rm pur}\cdot{\rm Acc}_{6h}$, the two quantities sensitive to the down/up assignment.

At the representative $Q_{\rm ST1}^{\rm avg}\ge -0.0889$ working point, both comparisons have the same calibrated background ratio, $N_{\rm bkg}/N_{\rm sig}=0.631$, and the same retained signal efficiency, $\epsilon_{\rm signal}^{B_{\rm ATL}+\mathcal{V}_{\rm ML}}=0.0845$.  With the learned ST2 classifier we find $\eta_\kappa=0.255$ and $P_{\rm pur}\cdot  {\rm Acc}_{6h}=0.218$; replacing ST2 by random down/up labels gives $\eta_\kappa=0.152$ and $P_{\rm pur}\cdot {\rm Acc}_{6h}=0.139$.  Thus, at fixed event selection and fixed signal efficiency, ST2 raises $\eta_\kappa$ by a factor $1.68$ and the purity-weighted six-parton exact fraction by a factor $1.57$.

As a direct reconstruction test, we also evaluate the spin-correlation coefficient $D$ used in standard top spin-correlation and entanglement analyses~\cite{Bernreuther:2015yna,ATLAS:2023fsd,CMS:2024pts}.  The spin analysis selection $\mathcal{V}_{\rm spin}$ requires a valid two step reconstruction and $300<m_{t\bar t}^{\rm reco}<430~{\rm GeV}$, in addition to the $B_{\rm ATL}$ baseline.  Here $m_{t\bar t}^{\rm reco}$ is formed from the six-jet candidate selected by ST1 and passed to ST2.  For comparison, the soft jet prescription selects the lower-energy light jet in each ST1 reconstructed top rest frame as the down-type candidate~\cite{Tweedie:2014yda}.  We form $D_{\rm soft}^{\rm sig+unm}$ with the same signal and unmatched normalization as $D_{\rm true}$; QCD is omitted from this assignment benchmark.  For the truth-level down-parton and ML-reconstructed down-jet directions, the averages below are taken over the selected events in the indicated component:
\begin{equation}\label{eq:Dtrue-DML}
\begin{aligned}
D_{\rm true}
&=
-3\,
\frac{
\left\langle
\hat n^{\rm truth}_{e,1}\cdot \hat n^{\rm truth}_{e,2}
\right\rangle_{\rm sig}
+
R_{\rm unm/sig}^{B_{\rm ATL}+\mathcal{V}_{\rm spin}}
\left\langle
\hat n^{\rm truth}_{e,1}\cdot \hat n^{\rm truth}_{e,2}
\right\rangle_{\rm unm}
}{
1+
R_{\rm unm/sig}^{B_{\rm ATL}+\mathcal{V}_{\rm spin}}
},\\
D_{\rm ML}
&=
-3\,
\frac{
\left\langle
\hat n^{\rm pred}_{e,1}\cdot \hat n^{\rm pred}_{e,2}
\right\rangle_{\rm sig}
+
R_{\rm unm/sig}^{B_{\rm ATL}+\mathcal{V}_{\rm spin}}
\left\langle
\hat n^{\rm pred}_{e,1}\cdot \hat n^{\rm pred}_{e,2}
\right\rangle_{\rm unm}
+
R_{\rm QCD/sig}^{B_{\rm ATL}+\mathcal{V}_{\rm spin}}
\left\langle
\hat n^{\rm pred}_{e,1}\cdot \hat n^{\rm pred}_{e,2}
\right\rangle_{\rm QCD}
}{
1+
R_{\rm unm/sig}^{B_{\rm ATL}+\mathcal{V}_{\rm spin}}
+
R_{\rm QCD/sig}^{B_{\rm ATL}+\mathcal{V}_{\rm spin}}
}.
\end{aligned}
\end{equation}

For the comparison below, $D_{\rm ML}^{\rm sig+unm}$ and $D_{\rm soft}^{\rm sig+unm}$ are evaluated using only the signal and unmatched components.

\begin{table}[t]
    \centering
    \small
    \setlength{\tabcolsep}{4pt}
    \resizebox{\linewidth}{!}{%
    \begin{tabular}{lcccccc}
        \hline
        Sampling method & Sampled sig/unm/QCD & $N_{\rm bkg}/N_{\rm sig}$ & $D_{\rm true}$ & $D_{\rm ML}^{\rm sig+unm}$ & $D_{\rm soft}^{\rm sig+unm}$ & $D_{\rm ML}$ \\
        \hline
        Calibrated component weights & $(148/2835/42)$ & $7.42$ & $-0.233\pm0.055$ & $-0.378\pm0.052$ & $0.157\pm0.052$ & $-0.439\pm0.143$ \\
        QCD limited event counts & $(7.63/14.61/42.00)$ & $7.42$ & $-0.227\pm0.375$ & $-0.380\pm0.360$ & $0.163\pm0.353$ & $-0.443\pm0.186$ \\
        \hline
    \end{tabular}
    }
    \caption{Spin-correlation results at $B_{\rm ATL}+\mathcal{V}_{\rm spin}$ from 4000 repeated samples.  The uncertainties are the standard deviations of the 4000 values.}
    \label{tab:spin-d-closure}
\end{table}

\begin{figure}
    \centering
    \includegraphics[width=0.48\linewidth]{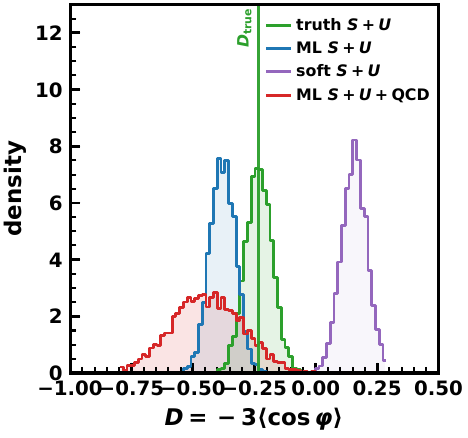}\hfill
    \includegraphics[width=0.48\linewidth]{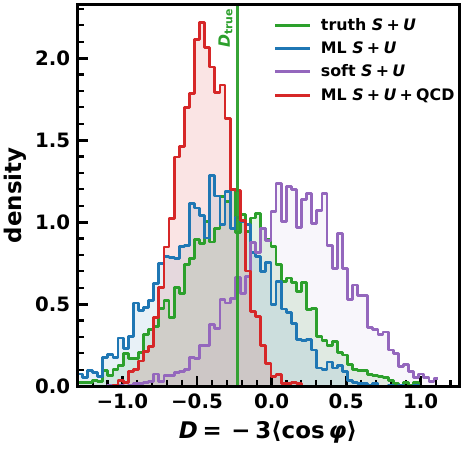}
    \caption{Distributions of the 4000 spin-correlation values at $B_{\rm ATL}+\mathcal{V}_{\rm spin}$.  The truth, ST2 MAP, and soft jet distributions are shown for signal plus unmatched events; the ST2 MAP result is also shown after including QCD.  The left panel uses calibrated component weights, while the right panel limits each sample to 42 QCD events and scales the signal and unmatched counts accordingly.}
    \label{fig:spin-d-bootstrap}
\end{figure}

\FloatBarrier
After applying $B_{\rm ATL}+\mathcal{V}_{\rm spin}$, the available pools contain 148 signal, 2835 unmatched $t\bar t$, and 42 QCD events.  We evaluate two sampling methods.  For the calibrated component weights, each pool is sampled with replacement at its available size, and the three component means are combined using the ratios in Eq.~\eqref{eq:Dtrue-DML}.  For the QCD limited event counts, every sample contains 42 QCD events and the same ratios set the mean signal and unmatched counts to 7.63 and 14.61.  The required integer counts fluctuate between 7 and 8 for signal and between 14 and 15 for unmatched.  Both methods use 4000 repeated samples.

The two sampling methods give consistent central values, while the QCD-limited event counts produce the larger statistical spread shown in Fig.~\ref{fig:spin-d-bootstrap}.  For the signal plus unmatched sample, the ST2 MAP result lies substantially closer to $D_{\rm true}$ than the soft jet reference: with calibrated component weights, the absolute differences are $0.145$ and $0.390$, respectively.

\section{Conclusion}
\label{sec:conclusion}

Fully hadronic $t\bar t$ events contain the largest branching fraction but are usually disfavoured for spin-correlation and entanglement studies because the down-type quarks from the two hadronic $W$ decays are not directly identified. Existing top-entanglement measurements and proposals have mostly relied on leptonic spin analyzers, while recent hadronic top-polarimetry studies have shown that the down-type direction can be statistically recovered from jet substructure information~\cite{ATLAS:2023fsd,CMS:2024pts,Dong:2024xsg}. This work addresses this reconstruction problem directly in the fully hadronic $t\bar t$ channel, where both down-type quarks must be inferred from jets.

We developed a two-stage jet-assignment network for this purpose.  The first classifier acts on all reconstructed jets and forms legal six-jet top-pair assignments using the per-jet logits together with a mass prior. The second classifier takes the selected six jets in the ordered topology $(b_1,q_{1a},q_{1b},b_2,q_{2a},q_{2b})$, enumerates the four possible down/up hypotheses, and predicts the final down/up assignment with the ST2 assignment score $S_2$. A conditional diffusion branch shares the event representation and supplies an auxiliary margin ranking signal during training. Its diffusion assignment scores are also retained for downstream cut variables.

The diffusion study shows that this auxiliary branch depends strongly on the loss. Standard denoising diffusion training, including the variants tested here, does not make the diffusion assignment score useful for ranking the four hypotheses. In contrast, the margin ranking objective directly optimizes the ordering of the true-assignment diffusion score relative to the wrong hypotheses, lifting the accuracy of $\arg\max_{\boldsymbol{\delta}}\mathcal{G}_{\boldsymbol{\delta}}$ above the random baseline. This supports the use of diffusion as an auxiliary training signal and as a source of additional cut variables.

The learned ST2 classifier gives a clear gain over a ST1 only benchmark with random down/up labels. At the representative $Q_{\rm ST1}^{\rm avg}\ge -0.0889$ working point with $\epsilon_{\rm signal}^{B_{\rm ATL}+\mathcal{V}_{\rm ML}}=0.0845$, both methods have the same calibrated background ratio, $N_{\rm bkg}/N_{\rm sig}=0.631$, but the learned ST2 classifier increases $\eta_\kappa$ from $0.152$ to $0.255$, a factor of $1.68$, and the purity-weighted six-parton exact fraction $P_{\rm pur}\cdot {\rm Acc}_{6h}$ from $0.139$ to $0.218$, a factor of $1.57$. These results indicate that down/up identification in the all-hadronic channel can retain useful spin-correlation information after realistic assignment ambiguities and calibrated QCD contamination are included.

For the spin-analysis selection, the calibrated component weights give $D_{\rm true}=-0.233\pm0.055$ and $D_{\rm ML}=-0.439\pm0.143$, while the QCD limited event counts give $D_{\rm true}=-0.227\pm0.375$ and $D_{\rm ML}=-0.443\pm0.186$.  On the signal plus unmatched sample, the ST2 MAP reconstruction is closer to the truth result than the soft jet reference.

\section{Acknowledgments}
The work of J.G. is supported by the Postdoctoral Fellowship Program (Grade C) of China Postdoctoral Science Foundation under Grant No. GZC20252775. We would like to thank Jia Liu, Xiao-Ping Wang, Chen Zhou and Congqiao Li for useful discussions.

\bibliographystyle{JHEP}
\bibliography{ref}

@InProceedings{Hang_2023_ICCV,
    author    = {Hang, Tiankai and Gu, Shuyang and Li, Chen and Bao, Jianmin and Chen, Dong and Hu, Han and Geng, Xin and Guo, Baining},
    title     = {Efficient Diffusion Training via Min-SNR Weighting Strategy},
    booktitle = {Proceedings of the IEEE/CVF International Conference on Computer Vision (ICCV)},
    month     = {October},
    year      = {2023},
    pages     = {7441-7451}
}

@inproceedings{
ho2021classifierfree,
title={Classifier-Free Diffusion Guidance},
author={Jonathan Ho and Tim Salimans},
booktitle={NeurIPS 2021 Workshop on Deep Generative Models and Downstream Applications},
year={2021},
url={https://openreview.net/forum?id=qw8AKxfYbI}
}

@inproceedings{
karras2022elucidating,
title={Elucidating the Design Space of Diffusion-Based Generative Models},
author={Tero Karras and Miika Aittala and Timo Aila and Samuli Laine},
booktitle={Advances in Neural Information Processing Systems},
editor={Alice H. Oh and Alekh Agarwal and Danielle Belgrave and Kyunghyun Cho},
year={2022},
url={https://openreview.net/forum?id=k7FuTOWMOc7}
}

@inproceedings{
salimans2022progressive,
title={Progressive Distillation for Fast Sampling of Diffusion Models},
author={Tim Salimans and Jonathan Ho},
booktitle={International Conference on Learning Representations},
year={2022},
url={https://openreview.net/forum?id=TIdIXIpzhoI}
}

@inproceedings{NEURIPS2020_4c5bcfec,
 author = {Ho, Jonathan and Jain, Ajay and Abbeel, Pieter},
 booktitle = {Advances in Neural Information Processing Systems},
 editor = {H. Larochelle and M. Ranzato and R. Hadsell and M.F. Balcan and H. Lin},
 pages = {6840--6851},
 publisher = {Curran Associates, Inc.},
 title = {Denoising Diffusion Probabilistic Models},
 url = {https://proceedings.neurips.cc/paper_files/paper/2020/file/4c5bcfec8584af0d967f1ab10179ca4b-Paper.pdf},
 volume = {33},
 year = {2020}
}

@inproceedings{
loshchilov2018decoupled,
title={Decoupled Weight Decay Regularization},
author={Ilya Loshchilov and Frank Hutter},
booktitle={International Conference on Learning Representations},
year={2019},
url={https://openreview.net/forum?id=Bkg6RiCqY7},
}

@InProceedings{pmlr-v139-nichol21a,
  title = 	 {Improved Denoising Diffusion Probabilistic Models},
  author =       {Nichol, Alexander Quinn and Dhariwal, Prafulla},
  booktitle = 	 {Proceedings of the 38th International Conference on Machine Learning},
  pages = 	 {8162--8171},
  year = 	 {2021},
  editor = 	 {Meila, Marina and Zhang, Tong},
  volume = 	 {139},
  series = 	 {Proceedings of Machine Learning Research},
  month = 	 {18--24 Jul},
  publisher =    {PMLR},
  url = 	 {https://proceedings.mlr.press/v139/nichol21a.html}
}

@InProceedings{Li_2023_ICCV,
    author    = {Li, Alexander C. and Prabhudesai, Mihir and Duggal, Shivam and Brown, Ellis and Pathak, Deepak},
    title     = {Your Diffusion Model is Secretly a Zero-Shot Classifier},
    booktitle = {Proceedings of the IEEE/CVF International Conference on Computer Vision (ICCV)},
    month     = {October},
    year      = {2023},
    pages     = {2206-2217}
}

@article{ATLAS:2020ccu,
    author = "Aad, Georges and others",
    collaboration = "ATLAS",
    title = "{Measurements of top-quark pair single- and double-differential cross-sections in the all-hadronic channel in $pp$ collisions at $\sqrt{s}=13~\textrm{TeV}$ using the ATLAS detector}",
    eprint = "2006.09274",
    archivePrefix = "arXiv",
    primaryClass = "hep-ex",
    reportNumber = "CERN-EP-2020-063",
    doi = "10.1007/JHEP01(2021)033",
    journal = "JHEP",
    volume = "01",
    pages = "033",
    year = "2021"
}

@article{ParticleDataGroup:2024cfk,
    author = "Navas, S. and others",
    collaboration = "Particle Data Group",
    title = "{Review of particle physics}",
    doi = "10.1103/PhysRevD.110.030001",
    journal = "Phys. Rev. D",
    volume = "110",
    number = "3",
    pages = "030001",
    year = "2024"
}

@article{Bernreuther:2008ju,
    author = "Bernreuther, Werner",
    title = "{Top quark physics at the LHC}",
    eprint = "0805.1333",
    archivePrefix = "arXiv",
    primaryClass = "hep-ph",
    reportNumber = "PITHA-08-09",
    doi = "10.1088/0954-3899/35/8/083001",
    journal = "J. Phys. G",
    volume = "35",
    pages = "083001",
    year = "2008"
}

@article{Bernreuther:2015yna,
    author = "Bernreuther, Werner and Heisler, Dennis and Si, Zong-Guo",
    title = "{A set of top quark spin correlation and polarization observables for the LHC: Standard Model predictions and new physics contributions}",
    eprint = "1508.05271",
    archivePrefix = "arXiv",
    primaryClass = "hep-ph",
    reportNumber = "TTK-15-16",
    doi = "10.1007/JHEP12(2015)026",
    journal = "JHEP",
    volume = "12",
    pages = "026",
    year = "2015"
}

@article{Tweedie:2014yda,
    author = "Tweedie, Brock",
    title = "{Better Hadronic Top Quark Polarimetry}",
    eprint = "1401.3021",
    archivePrefix = "arXiv",
    primaryClass = "hep-ph",
    reportNumber = "PITT-PACC-1315",
    doi = "10.1103/PhysRevD.90.094010",
    journal = "Phys. Rev. D",
    volume = "90",
    number = "9",
    pages = "094010",
    year = "2014"
}

@article{Dong:2024xsg,
    author = "Dong, Zhongtian and Gon{\c{c}}alves, Dorival and Kong, Kyoungchul and Larkoski, Andrew J. and Navarro, Alberto",
    title = "{Hadronic top quark polarimetry with ParticleNet}",
    eprint = "2407.01663",
    archivePrefix = "arXiv",
    primaryClass = "hep-ph",
    doi = "10.1016/j.physletb.2025.139314",
    journal = "Phys. Lett. B",
    volume = "862",
    pages = "139314",
    year = "2025"
}

@article{Dong:2024xsb,
    author = "Dong, Zhongtian and Gon{\c{c}}alves, Dorival and Kong, Kyoungchul and Larkoski, Andrew J. and Navarro, Alberto",
    title = "{Analytical insights on hadronic top quark polarimetry}",
    eprint = "2407.07147",
    archivePrefix = "arXiv",
    primaryClass = "hep-ph",
    doi = "10.1007/JHEP02(2025)117",
    journal = "JHEP",
    volume = "02",
    pages = "117",
    year = "2025"
}

@article{ATLAS:2023fsd,
    author = "Aad, Georges and others",
    collaboration = "ATLAS",
    title = "{Observation of quantum entanglement with top quarks at the ATLAS detector}",
    eprint = "2311.07288",
    archivePrefix = "arXiv",
    primaryClass = "hep-ex",
    reportNumber = "CERN-EP-2023-230",
    doi = "10.1038/s41586-024-07824-z",
    journal = "Nature",
    volume = "633",
    number = "8030",
    pages = "542--547",
    year = "2024"
}

@article{CMS:2024pts,
    author = "Hayrapetyan, Aram and others",
    collaboration = "CMS",
    title = "{Observation of quantum entanglement in top quark pair production in proton{\textendash}proton collisions at $\sqrt{s} = 13$ TeV}",
    eprint = "2406.03976",
    archivePrefix = "arXiv",
    primaryClass = "hep-ex",
    reportNumber = "CMS-TOP-23-001, CERN-EP-2024-137",
    doi = "10.1088/1361-6633/ad7e4d",
    journal = "Rept. Prog. Phys.",
    volume = "87",
    number = "11",
    pages = "117801",
    year = "2024"
}

@article{Fenton:2020woz,
    author = "Fenton, Michael James and Shmakov, Alexander and Ho, Ta-Wei and Hsu, Shih-Chieh and Whiteson, Daniel and Baldi, Pierre",
    title = "{Permutationless many-jet event reconstruction with symmetry preserving attention networks}",
    eprint = "2010.09206",
    archivePrefix = "arXiv",
    primaryClass = "hep-ex",
    doi = "10.1103/PhysRevD.105.112008",
    journal = "Phys. Rev. D",
    volume = "105",
    number = "11",
    pages = "112008",
    year = "2022"
}

@article{Alwall:2014hca,
    author = "Alwall, J. and Frederix, R. and Frixione, S. and Hirschi, V. and Maltoni, F. and Mattelaer, O. and Shao, H. -S. and Stelzer, T. and Torrielli, P. and Zaro, M.",
    title = "{The automated computation of tree-level and next-to-leading order differential cross sections, and their matching to parton shower simulations}",
    eprint = "1405.0301",
    archivePrefix = "arXiv",
    primaryClass = "hep-ph",
    reportNumber = "CERN-PH-TH-2014-064, CP3-14-18, LPN14-066, MCNET-14-09, ZU-TH-14-14",
    doi = "10.1007/JHEP07(2014)079",
    journal = "JHEP",
    volume = "07",
    pages = "079",
    year = "2014"
}

@article{Artoisenet:2012st,
    author = "Artoisenet, Pierre and Frederix, Rikkert and Mattelaer, Olivier and Rietkerk, Robbert",
    title = "{Automatic spin-entangled decays of heavy resonances in Monte Carlo simulations}",
    eprint = "1212.3460",
    archivePrefix = "arXiv",
    primaryClass = "hep-ph",
    reportNumber = "NIKHEF-2012-021, CERN-PH-TH-2012-329",
    doi = "10.1007/JHEP03(2013)015",
    journal = "JHEP",
    volume = "03",
    pages = "015",
    year = "2013"
}

@article{Bierlich:2022pfr,
    author = "Bierlich, Christian and others",
    title = "{A comprehensive guide to the physics and usage of PYTHIA 8.3}",
    eprint = "2203.11601",
    archivePrefix = "arXiv",
    primaryClass = "hep-ph",
    reportNumber = "LU-TP 22-16, MCNET-22-04, FERMILAB-PUB-22-227-SCD",
    doi = "10.21468/SciPostPhysCodeb.8",
    journal = "SciPost Phys. Codeb.",
    volume = "2022",
    pages = "8",
    year = "2022"
}

@article{deFavereau:2013fsa,
    author = "de Favereau, J. and Delaere, C. and Demin, P. and Giammanco, A. and Lema{\^\i}tre, V. and Mertens, A. and Selvaggi, M.",
    collaboration = "DELPHES 3",
    title = "{DELPHES 3, A modular framework for fast simulation of a generic collider experiment}",
    eprint = "1307.6346",
    archivePrefix = "arXiv",
    primaryClass = "hep-ex",
    doi = "10.1007/JHEP02(2014)057",
    journal = "JHEP",
    volume = "02",
    pages = "057",
    year = "2014"
}

@article{Lee:2020qil,
    author = "Lee, Jason Sang Hun and Park, Inkyu and Watson, Ian James and Yang, Seungjin",
    title = "{Zero-permutation jet-parton assignment using a self-attention network}",
    eprint = "2012.03542",
    archivePrefix = "arXiv",
    primaryClass = "hep-ex",
    doi = "10.1007/s40042-024-01037-3",
    journal = "J. Korean Phys. Soc.",
    volume = "84",
    number = "6",
    pages = "427--438",
    year = "2024"
}

@article{CMS:2018tye,
    author = "Sirunyan, Albert M and others",
    collaboration = "CMS",
    title = "{Measurement of the top quark mass in the all-jets final state at $\sqrt{s} =$ 13 TeV and combination with the lepton+jets channel}",
    eprint = "1812.10534",
    archivePrefix = "arXiv",
    primaryClass = "hep-ex",
    reportNumber = "CMS-TOP-17-008, CERN-EP-2018-310",
    doi = "10.1140/epjc/s10052-019-6788-2",
    journal = "Eur. Phys. J. C",
    volume = "79",
    number = "4",
    pages = "313",
    year = "2019"
}

@article{Ba:2016jcy,
    author = "Ba, Jimmy Lei and Kiros, Jamie Ryan and Hinton, Geoffrey E.",
    title = "{Layer Normalization}",
    eprint = "1607.06450",
    archivePrefix = "arXiv",
    primaryClass = "stat.ML",
    month = "7",
    year = "2016"
}

@inproceedings{Nair2010RectifiedLU,
  title={Rectified Linear Units Improve Restricted Boltzmann Machines},
  author={Vinod Nair and Geoffrey E. Hinton},
  booktitle={International Conference on Machine Learning},
  year={2010},
  url={https://api.semanticscholar.org/CorpusID:15539264}
}

@article{Qu:2019gqs,
    author = "Qu, Huilin and Gouskos, Loukas",
    title = "{ParticleNet: Jet Tagging via Particle Clouds}",
    eprint = "1902.08570",
    archivePrefix = "arXiv",
    primaryClass = "hep-ph",
    doi = "10.1103/PhysRevD.101.056019",
    journal = "Phys. Rev. D",
    volume = "101",
    number = "5",
    pages = "056019",
    year = "2020"
}

@phdthesis{Poggi:2021thesis,
    author = "Poggi, Riccardo",
    title = "{Top-quark pair production cross-section measurements in the all-hadronic decay channel at the ATLAS experiment and hardware-based track reconstruction for the ATLAS trigger HL-LHC upgrade}",
    school = "University of Geneva",
    year = "2021",
    doi = "10.13097/archive-ouverte/unige:151144",
    url = "https://archive-ouverte.unige.ch/unige:151144"
}

@article{Afik:2020onf,
    author = "Afik, Yoav and de Nova, Juan Ram{\'o}n Mu{\~n}oz",
    title = "{Entanglement and quantum tomography with top quarks at the LHC}",
    eprint = "2003.02280",
    archivePrefix = "arXiv",
    primaryClass = "quant-ph",
    doi = "10.1140/epjp/s13360-021-01902-1",
    journal = "Eur. Phys. J. Plus",
    volume = "136",
    number = "9",
    pages = "907",
    year = "2021"
}

@article{Severi:2021cnj,
    author = "Severi, Claudio and Boschi, Cristian Degli Esposti and Maltoni, Fabio and Sioli, Maximiliano",
    title = "{Quantum tops at the LHC: from entanglement to Bell inequalities}",
    eprint = "2110.10112",
    archivePrefix = "arXiv",
    primaryClass = "hep-ph",
    doi = "10.1140/epjc/s10052-022-10245-9",
    journal = "Eur. Phys. J. C",
    volume = "82",
    number = "4",
    pages = "285",
    year = "2022"
}

@article{Aguilar-Saavedra:2022uye,
    author = "Aguilar-Saavedra, J. A. and Casas, J. A.",
    title = "{Improved tests of entanglement and Bell inequalities with LHC tops}",
    eprint = "2205.00542",
    archivePrefix = "arXiv",
    primaryClass = "hep-ph",
    reportNumber = "IFT-UAM/CSIC-22-45",
    doi = "10.1140/epjc/s10052-022-10630-4",
    journal = "Eur. Phys. J. C",
    volume = "82",
    number = "8",
    pages = "666",
    year = "2022"
}

@article{Han:2023fci,
    author = "Han, Tao and Low, Matthew and Wu, Tong Arthur",
    title = "{Quantum entanglement and Bell inequality violation in semi-leptonic top decays}",
    eprint = "2310.17696",
    archivePrefix = "arXiv",
    primaryClass = "hep-ph",
    reportNumber = "PITT-PACC-2316",
    doi = "10.1007/JHEP07(2024)192",
    journal = "JHEP",
    volume = "07",
    pages = "192",
    year = "2024"
}

@article{CMS:2024zkc,
    author = "Hayrapetyan, Aram and others",
    collaboration = "CMS",
    title = "{Measurements of polarization and spin correlation and observation of entanglement in top quark pairs using lepton+jets events from proton--proton collisions at $\sqrt{s}=13$ TeV}",
    eprint = "2409.11067",
    archivePrefix = "arXiv",
    primaryClass = "hep-ex",
    reportNumber = "CMS-TOP-23-007, CERN-EP-2024-231",
    doi = "10.1103/PhysRevD.110.112016",
    journal = "Phys. Rev. D",
    volume = "110",
    number = "11",
    pages = "112016",
    year = "2024"
}

@article{Brandenburg:2002xr,
    author = "Brandenburg, Arnd and Si, Z. G. and Uwer, P.",
    title = "{QCD corrected spin analyzing power of jets in decays of polarized top quarks}",
    eprint = "hep-ph/0205023",
    archivePrefix = "arXiv",
    primaryClass = "hep-ph",
    reportNumber = "PITHA-02-07, DESY-02-055, TTP02-04",
    doi = "10.1016/S0370-2693(02)02098-1",
    journal = "Phys. Lett. B",
    volume = "539",
    pages = "235--241",
    year = "2002"
}

@article{Fraser:2018ieu,
    author = "Fraser, Katherine and Schwartz, Matthew D.",
    title = "{Jet Charge and Machine Learning}",
    eprint = "1803.08066",
    archivePrefix = "arXiv",
    primaryClass = "hep-ph",
    doi = "10.1007/JHEP10(2018)093",
    journal = "JHEP",
    volume = "10",
    pages = "093",
    year = "2018"
}

@article{Erdmann:2013rxa,
    author = "Erdmann, Johannes and Guindon, Stefan and Kroeninger, Kevin and Lemmer, Boris and Nackenhorst, Olaf and Quadt, Arnulf and Stolte, Philipp",
    title = "{A likelihood-based reconstruction algorithm for top-quark pairs and the KLFitter framework}",
    eprint = "1312.5595",
    archivePrefix = "arXiv",
    primaryClass = "hep-ex",
    doi = "10.1016/j.nima.2014.02.029",
    journal = "Nucl. Instrum. Meth. A",
    volume = "748",
    pages = "18--25",
    year = "2014"
}

@article{Erdmann:2019evj,
    author = "Erdmann, Johannes and Kallage, Tim and Kr{\"o}ninger, Kevin and Nackenhorst, Olaf",
    title = "{From the bottom to the top---reconstruction of $t\bar{t}$ events with deep learning}",
    eprint = "1907.11181",
    archivePrefix = "arXiv",
    primaryClass = "hep-ex",
    doi = "10.1088/1748-0221/14/11/P11015",
    journal = "JINST",
    volume = "14",
    number = "11",
    pages = "P11015",
    year = "2019"
}

@article{Shmakov:2021qdz,
    author = "Shmakov, Alexander and Fenton, Michael James and Ho, Ta-Wei and Hsu, Shih-Chieh and Whiteson, Daniel and Baldi, Pierre",
    title = "{SPANet: Generalized permutationless set assignment for particle physics using symmetry preserving attention}",
    eprint = "2106.03898",
    archivePrefix = "arXiv",
    primaryClass = "hep-ex",
    doi = "10.21468/SciPostPhys.12.5.178",
    journal = "SciPost Phys.",
    volume = "12",
    number = "5",
    pages = "178",
    year = "2022"
}

@article{Ehrke:2023cpn,
    author = "Ehrke, Lukas and Raine, John Andrew and Zoch, Knut and Guth, Manuel and Golling, Tobias",
    title = "{Topological reconstruction of particle physics processes using graph neural networks}",
    eprint = "2303.13937",
    archivePrefix = "arXiv",
    primaryClass = "hep-ph",
    doi = "10.1103/PhysRevD.107.116019",
    journal = "Phys. Rev. D",
    volume = "107",
    number = "11",
    pages = "116019",
    year = "2023"
}

@article{Birch-Sykes:2024gij,
    author = "Birch-Sykes, Callum and Le, Brian and Peters, Yvonne and Simpson, Ethan and Zhang, Zihan",
    title = "{Reconstructing short-lived particles using hypergraph representation learning}",
    eprint = "2402.10149",
    archivePrefix = "arXiv",
    primaryClass = "hep-ph",
    doi = "10.1103/PhysRevD.111.032004",
    journal = "Phys. Rev. D",
    volume = "111",
    number = "3",
    pages = "032004",
    year = "2025"
}

@article{Mahlon:2010gw,
    author = "Mahlon, Gregory and Parke, Stephen J.",
    title = {Spin Correlation Effects in Top Quark Pair Production at the LHC},
    eprint = "1001.3422",
    archivePrefix = "arXiv",
    primaryClass = "hep-ph",
    reportNumber = "FERMILAB-PUB-09-662-T",
    doi = "10.1103/PhysRevD.81.074024",
    journal = "Phys. Rev. D",
    volume = "81",
    pages = "074024",
    year = "2010",
}

@article{Krohn:2009wm,
    author = "Krohn, David and Shelton, Jessie and Wang, Lian-Tao",
    title = {Measuring the Polarization of Boosted Hadronic Tops},
    eprint = "0909.3855",
    archivePrefix = "arXiv",
    primaryClass = "hep-ph",
    doi = "10.1007/JHEP07(2010)041",
    journal = "JHEP",
    volume = "07",
    pages = "041",
    year = "2010",
}

@article{Fabbrichesi:2021npl,
    author = "Fabbrichesi, M. and Floreanini, R. and Panizzo, G.",
    title = {Testing Bell Inequalities at the LHC with Top-Quark Pairs},
    eprint = "2102.11883",
    archivePrefix = "arXiv",
    primaryClass = "hep-ph",
    doi = "10.1103/PhysRevLett.127.161801",
    journal = "Phys. Rev. Lett.",
    volume = "127",
    number = "16",
    pages = "161801",
    year = "2021",
}

@article{Afik:2022kwm,
    author = "Afik, Yoav and de Nova, Juan Ram{\'o}n Mu{\~n}oz",
    title = {Quantum information with top quarks in QCD},
    eprint = "2203.05582",
    archivePrefix = "arXiv",
    primaryClass = "quant-ph",
    doi = "10.22331/q-2022-09-29-820",
    journal = "Quantum",
    volume = "6",
    pages = "820",
    year = "2022",
}

@article{Qiu:2022xvr,
    author = "Qiu, Shikai and Han, Shuo and Ju, Xiangyang and Nachman, Benjamin and Wang, Haichen",
    title = {Holistic approach to predicting top quark kinematic properties with the covariant particle transformer},
    eprint = "2203.05687",
    archivePrefix = "arXiv",
    primaryClass = "hep-ph",
    doi = "10.1103/PhysRevD.107.114029",
    journal = "Phys. Rev. D",
    volume = "107",
    number = "11",
    pages = "114029",
    year = "2023",
}

@article{Dillon:2025dxr,
    author = "Dillon, Barry M. and Spannowsky, Michael",
    title = {Theory-informed neural networks for particle physics},
    eprint = "2507.13447",
    archivePrefix = "arXiv",
    primaryClass = "hep-ph",
    reportNumber = "IPPP/25/47",
    doi = "10.1088/2632-2153/ae47ba",
    journal = "Mach. Learn. Sci. Tech.",
    volume = "7",
    number = "2",
    pages = "025010",
    year = "2026",
}

@article{Barman:2024wfx,
    author = "Barman, Rahool Kumar and Biswas, Sumit",
    title = "{Top-philic machine learning}",
    eprint = "2407.00183",
    archivePrefix = "arXiv",
    primaryClass = "hep-ph",
    doi = "10.1140/epjs/s11734-024-01237-9",
    journal = "Eur. Phys. J. ST",
    volume = "233",
    number = "15-16",
    pages = "2497--2530",
    year = "2024"
}

\end{document}